\documentclass[a4paper,fontsize=11pt,DIV=12,twocolumn=false]{scrartcl}

\usepackage[utf8]{inputenc}  
\usepackage[T1]{fontenc}  
\usepackage{lmodern}  
\usepackage{array}  
\usepackage{graphicx}  
\usepackage{float}  
\usepackage{amsmath}  
\usepackage{amssymb}  
\usepackage{amsfonts}  
\usepackage{textcomp}  
\usepackage{threeparttable}
\usepackage{enumitem}  
\usepackage{algpseudocode}  
\usepackage{algorithm}  
\usepackage{longtable}
\usepackage{pdflscape}
\usepackage{microtype}
\usepackage[labelfont=bf,font=small,format=plain,indention=0cm]{caption}  
\usepackage{authblk}  

\usepackage{booktabs}  
\usepackage{multirow}  

\usepackage[
    group-separator={\,},
    separate-uncertainty=true,
    bracket-ambiguous-numbers=false,
    retain-zero-uncertainty=true,
    multi-part-units=single,
    list-units=single,
    range-units=single,
]{siunitx}  
\DeclareSIUnit[per-mode=symbol]{\gpcc}{\gram\per\cubic\centi\metre}  
\DeclareSIUnit{\bar}{bar}  
\DeclareSIUnit{\angstrom}{\text{Å}}  
\DeclareSIUnit{\atm}{\text{atm}}  
\DeclareSIUnit[per-mode=symbol]{\gpmol}{\gram\per\mole}  
\DeclareSIUnit[per-mode=symbol]{\ev}{\electronvolt}  
\DeclareSIUnit[per-mode=symbol]{\evpa}{\electronvolt\per\angstrom}  
\DeclareSIUnit[per-mode=symbol]{\mevpa}{\milli\electronvolt\per\angstrom}  
\DeclareSIUnit[per-mode=symbol]{\evpac}{\electronvolt\per\cubic\angstrom}  
\DeclareSIUnit[per-mode=symbol]{\mevpac}{\milli\electronvolt\per\cubic\angstrom}  
\DeclareSIUnit[per-mode=symbol]{\nspday}{\nano\second\per\day}  

\DeclareMathOperator{\Unif}{Unif}
\DeclareMathOperator{\LogUnif}{LogUnif}
\usepackage[
    backend=biber,
    style=nature,
    sorting=none,
    maxnames=99,
    maxcitenames=2,
    maxbibnames=5  
]{biblatex}
\usepackage[hidelinks]{hyperref}  

\renewbibmacro*{journal}{%
  \iffieldundef{shortjournal}
    {%
      \iffieldundef{journaltitle}
        {}
        {%
          \printtext[journaltitle]
            {%
              \printfield[titlecase]{journaltitle}%
              \setunit{\subtitlepunct}%
              \printfield[titlecase]{journalsubtitle}%
             }%
         }%
    }
    {\printtext[journaltitle]{\printfield[titlecase]{shortjournal}}}%
}
\AtEveryBibitem{%
  \clearfield{day}%
  \clearfield{month}%
  \clearfield{endday}%
  \clearfield{endmonth}%
  \clearfield{issue}%
  \clearfield{eprintclass}%
}
\renewbibmacro*{title}{%
  \printfield[titlecase]{title}%
  \addperiod}
\DeclareBibliographyDriver{misc}{%
  \usebibmacro{author}%
  \setunit{\labelnamepunct}\newblock
  \usebibmacro{title}%
  \newunit\newblock
  \iffieldundef{eprint}
    {}
    {\printtext{Preprint at \url{https://arxiv.org/abs/\thefield{eprint}}}}%
  \setunit{\addspace}
  \printtext[parens]{\printdate}%
  \finentry}

\newcommand{\tg}{T\ensuremath{_\mathrm{g}}}
\newcommand{\tgfit}{T\ensuremath{_\mathrm{g}^\mathrm{fit}}}

\usepackage{amsmath,amsfonts,bm}

\newcommand{\MLP}{\text{MLP}}
\newcommand{\medium}{\text{M}}
\newcommand{\short}{\text{S}}

\newcommand{\rcut}{r_\text{c}}
\newcommand{\fcut}{\eta_\text{c}}
\newcommand{\rcutmod}{r_\text{c}^\text{mod}}
\newcommand{\rcutshort}{r_\text{c}^\text{S}}
\newcommand{\rcutmid}{r_\text{c}^\text{M}}

\newcommand{\brho}{\boldsymbol{\rho}}

\newcommand{\brij}{\boldsymbol{r}_{ij}}

\newcommand{\rij}{{r}_{ij}}

\newcommand{\bx}{\boldsymbol{x}}

\newcommand{\ngpu}{n_\text{gpu}}
\newcommand{\natom}{n_\text{atom}}
\newcommand{\nchn}{n_\text{channel}}
\newcommand{\nbasis}{n_\text{basis}}
\newcommand{\nfeat}{n_\text{feat.}}

\newcommand{\localenv}{\mathcal{N}}
\newcommand{\fullenv}{\mathcal{N}}
\newcommand{\localshort}{\mathcal{N}_\text{S}}
\newcommand{\localmid}{\mathcal{N}_\text{M}}

\newcommand{\nedge}{{n_\text{edge}}}
\newcommand{\nlayer}{{n_\text{layer}}}
\newcommand{\nprelayer}{{n_\text{pre}}}
\newcommand{\npostlayer}{{n_\text{post}}}
\newcommand{\nsub}{{n_\text{sub}}}
\newcommand{\mlpd}{{n_\text{MLP-layer}}}
\newcommand{\rmax}{r_\text{max}}
\newcommand{\rmod}{r_\text{mod}}

\newcommand{\lmax}{L_\text{max}}

\newcommand{\concat}{\mathbin\Vert}

\def\1{\bm{1}}

\def\rva{{\mathbf{a}}}
\def\rvb{{\mathbf{b}}}
\def\rvc{{\mathbf{c}}}

\def\rvt{{\mathbf{t}}}

\def\txi{{\tilde{\bm{\xi}}}}
\def\vxi{{\bm{\xi}}}
\def\valpha{{\bm{\alpha}}}
\def\vrho{{\bm{\rho}}}

\def\va{{\bm{a}}}
\def\vb{{\bm{b}}}
\def\vc{{\bm{c}}}

\def\ve{{\bm{e}}}
\def\vf{{\bm{f}}}

\def\vh{{\bm{h}}}

\def\vk{{\bm{k}}}

\def\vq{{\bm{q}}}
\def\vr{{\bm{r}}}

\def\vt{{\bm{t}}}

\def\vv{{\bm{v}}}
\def\vw{{\bm{w}}}

\def\mA{{\bm{A}}}
\def\mB{{\bm{B}}}

\def\mQ{{\bm{Q}}}

\def\mU{{\bm{U}}}
\def\mV{{\bm{V}}}
\def\mW{{\bm{W}}}
\def\mX{{\bm{X}}}
\def\mY{{\bm{Y}}}
\def\mZ{{\bm{Z}}}

\def\msigma{{\bm{\sigma}}}
\def\mSigma{{\bm{\Sigma}}}
\def\mvarepsilon{\bm{\varepsilon}}

\DeclareMathAlphabet{\mathsfit}{\encodingdefault}{\sfdefault}{m}{sl}
\SetMathAlphabet{\mathsfit}{bold}{\encodingdefault}{\sfdefault}{bx}{n}

\def\sR{{\mathbb{R}}}

\def\sZ{{\mathbb{Z}}}

\newcolumntype{x}[1]{>{\centering\let\newline\\\arraybackslash\hspace{0pt}}p{#1}}

\title{SimPoly: Simulation of Polymers with Machine Learning Force Fields Derived from \\ First Principles}

\author[1]{Gregor N.\ C.\ Simm%
\thanks{These authors contributed equally to this work.}%
\thanks{Correspondence to \texttt{gregorsimm@microsoft.com} and  \texttt{lixinsun@microsoft.com}}}
\author[1]{Jean H\'{e}lie\textsuperscript{$\ast$}}
\author[1]{Hannes Schulz}
\author[1]{Yicheng Chen}
\author[1]{Anna Kuzina}
\author[1]{Ernesto Martinez-Baez}
\author[2]{Guillem Simeon}
\author[2]{Piero Gasparotto}
\author[2]{Gabriele Tocci}
\author[2]{Chi Chen}
\author[1]{Yatao Li}
\author[1]{Lixue Cheng}
\author[1]{Zun Wang}
\author[1]{Bichlien H.\ Nguyen}
\author[1]{Jake A.\ Smith}
\author[1]{Lixin Sun\textsuperscript{$\ast$\,\textdagger}}%

\affil[1]{Microsoft Research, AI for Science}
\affil[2]{Microsoft Quantum}

\date{\today}

\begin{document}

\maketitle


\begin{abstract}
    Polymers are a versatile class of materials with widespread industrial applications.
    Advanced computational tools could revolutionize their design, but their complex, multi-scale nature poses significant modeling challenges.
    Conventional force fields often lack the accuracy and transferability required to capture the intricate interactions governing polymer behavior.
    Conversely, quantum-chemical methods are computationally prohibitive for the large systems and long timescales required to simulate relevant polymer phenomena.
    Here, we overcome these limitations with a machine learning force field (MLFF) approach.
    We demonstrate that macroscopic properties for a broad range of polymers can be predicted \textit{ab initio}, without fitting to experimental data.
    Specifically, we develop a fast and scalable MLFF to accurately predict polymer densities, outperforming established classical force fields.
    Our MLFF also captures second-order phase transitions, enabling the prediction of glass transition temperatures.
    To accelerate progress in this domain, we introduce a benchmark of experimental bulk properties for \num{130} polymers and an accompanying quantum-chemical dataset.
    This work lays the foundation for a fully \textit{in silico} design pipeline for next-generation polymeric materials.
\end{abstract}

\section{Introduction}

Polymeric materials are foundational to modern life, embedded in everything from the clothes we wear and the food we consume to high-performance materials in aerospace, electronics, and medicine.
The ubiquity of polymers presents a significant opportunity for impact through improved tools for their functional design and a deeper understanding of their environmental and health effects\supercite{Mohanty2022Sustainable}.
The behavior of polymeric systems is complex, spanning multiple length and time scales.
This behavior arises from a combination of diverse local interactions within monomer structures and long-range interactions between polymer chains, which together determine bulk properties and processing behaviors.
These complexities demand specialized modeling strategies that can capture both the molecular detail and macroscopic behavior of polymer systems\supercite{Gartner2019Modeling,Schmid2023Understanding}.

Computationally efficient classical force fields (FFs) suitable for large-scale molecular dynamics (MD) simulations are currently the backbone of molecular modeling of polymers\supercite{Gartner2019Modeling,Li2015Molecular}.
These FFs describe interatomic interactions via a series of parametrized energy terms designed to capture covalent and non-covalent (e.g., van der Waals, electrostatic) interactions.
Through a time-consuming process, experts fit the parameters of these models to a combination of computational and experimental data\supercite{Brooks1983CHARMM,Mayo1990DREIDING,Sun1994PCFF,Sun1998COMPASS}.
A critical drawback of classical FFs is their limited transferability, i.e., their inability to accurately simulate conditions beyond those for which they were optimized.
While classical FFs are successful in specialized domains with limited chemical diversity (e.g., proteins in aqueous solution), the vast chemical space of synthetic polymers hinders their applicability due to a lack of transferability.
In addition, they are unable to model chemical reactions, as they rely on fixed bonding topologies.
For instance, the synthesis and degradation of polymers involve complex chemical transformations that cannot be captured by classical FFs.
Reactive FFs such as ReaxFF\supercite{vanDuin2001ReaxFF} have been developed to address this limitation; however, they often require laborious reparameterization\supercite{DeAngelis2024Enhancing,Zhang2018Improvement} as their accuracy is highly system-dependent\supercite{Qi2013Comparison,Bertels2020Benchmarking}.

Machine learning force fields (MLFFs) are trained on quantum-chemical data to predict energies and related quantities at a fraction of the computational cost\supercite{Behler2017First,Unke2021MLFFs,Deringer2019MLIPs}.
Their applications range from high-pressure systems\supercite{Cheng2020Evidence,Deringer2021Origins,yang2024mattersim} to drug design\supercite{Takaba2024Machinelearned} and the discovery of molecular reaction mechanisms\supercite{Zhang2024Exploring}.
MLFFs have the potential to revolutionize atomistic simulations of polymers by overcoming the fundamental limitations of classical FFs.
First, so-called universal or foundational MLFFs have demonstrated transferability across diverse chemical systems\supercite{Batatia2024Foundation,yang2024mattersim,Wood2025UMA}.
Second, they can inherently model bond-breaking and bond-forming transformations without extensive reparameterization.
Finally, unlike classical FFs, MLFFs can be systematically improved with more training data to accurately describe the complex interactions in the vast chemical space of polymers.

Despite their promise, MLFFs still face significant methodological challenges.
For example, they are computationally more expensive than classical FFs, which hinders their application to large systems and long-timescale MD simulations.
However, progress in MLFF development has so far been driven by benchmarks focused on reproducing quantum-chemical reference data rather than experimental validation\supercite{Ramakrishnan2014,Chanussot2021OC20,Tran2023OC22,Levine2025OMol25,Gharakhanyan2025OMC25,Riebesell2025Framework}.
Recent studies have shown that strong performance on computational benchmarks does not always translate to reliable predictions of experimental outcomes\supercite{Poltavsky2025CrashA,Poltavsky2025CrashB,Mannan2025Evaluating,Han2025Benchmarking}.
This highlights the need for robust validation against experimental data.

MLFF development has largely focused on crystalline and small-molecule benchmarks, leaving polymer systems relatively unexplored\supercite{Long2024Polymers}.
Recently, \citeauthor{Matsumura2025Generator} trained an MLFF on system-specific quantum-chemical data to reproduce the bulk properties of individual polymers\supercite{Matsumura2025Generator}.
However, the model's ability to describe multiple polymers simultaneously or generalize to new ones was not assessed.

\begin{figure}[H]
    \centering
    \includegraphics[width=\textwidth]{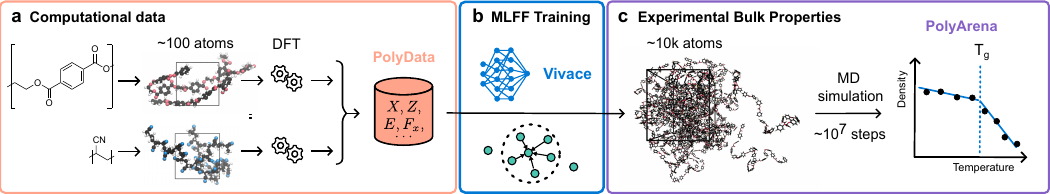}
    \caption{
        Prediction of experimental bulk properties for a wide range of polymers through MD simulations with a MLFF trained solely on \textit{ab initio} data.
        \textbf{a}, We generate a targeted quantum-chemical dataset containing small, atomistic polymeric model systems covering the complex intra- and inter-molecular interactions characteristic of polymers called PolyData.
        \textbf{b}, We introduce Vivace, a fast and scalable, SE(3)-equivariant MLFF architecture optimized for large-scale MD simulations, and train it on PolyData, together with a collection of other public data sets.
        \textbf{c}, We run MD simulations driven by Vivace using large model systems of polymers to predict experimentally measured (volumetric mass) densities. We also observe second-order phase transitions, enabling the determination of glass transition temperatures, \tg s.
        To systematically evaluate the performance of MLFFs on this challenging task, we introduce PolyArena, a benchmark of experimental bulk properties for \num{130} polymers.
    }
    \label{fig:intro}
\end{figure}

This work makes several contributions (Fig.~\ref{fig:intro}) to advance the atomistic modeling of polymers with MLFFs.
First, we develop Vivace, a strictly local MLFF architecture tailored for the large-scale atomistic simulations required for polymer modeling.
Second, we introduce PolyArena, a benchmark for evaluating MLFFs on experimentally measured polymer properties, namely (volumetric mass) densities and glass transition temperatures.
Third, we present PolyData, an accompanying dataset specifically designed for training MLFFs on these systems.
Through MD simulations driven by Vivace, we accurately predict the densities of polymers, outperforming two classical FFs and matching or exceeding the performance of two existing MLFFs.
Finally, we demonstrate that Vivace can capture second-order phase transitions, enabling the estimation of polymer glass transition temperatures.

\section{Results}

\subsection{Experimental Benchmarks and Computational Data for Polymer Simulations}

\begin{figure}[H]
    \centering
    \includegraphics[width=\textwidth]{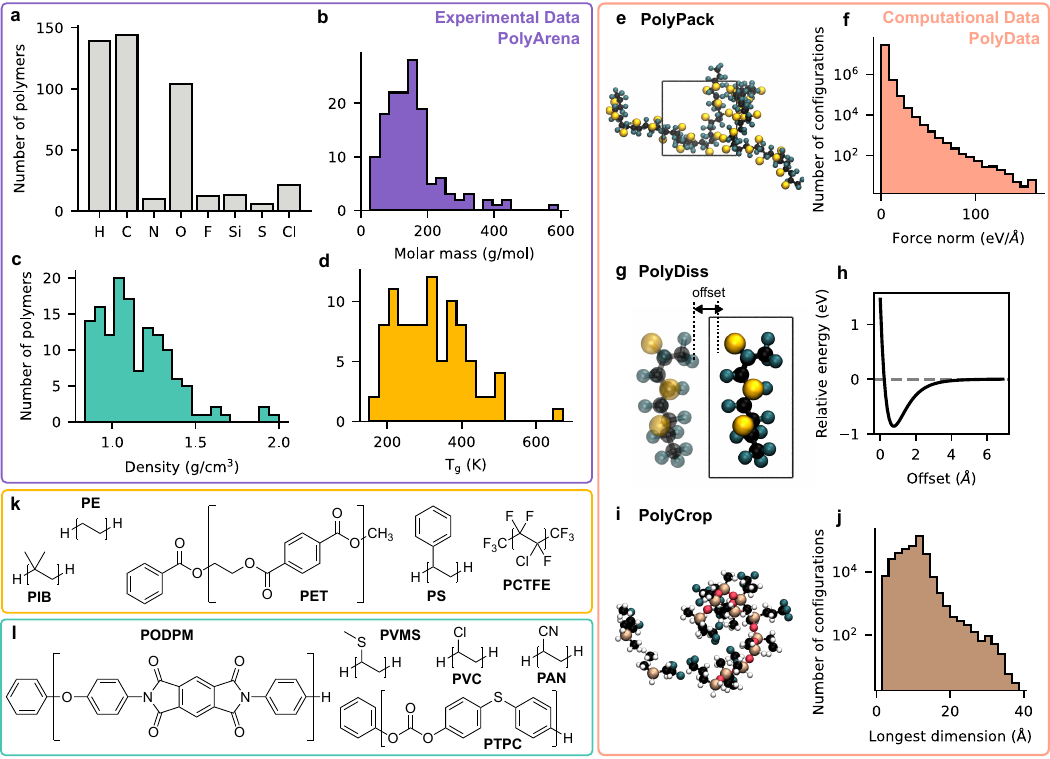}
    \caption{
        Overview of the experimental and computational data presented in this study.
        PolyArena (a--d) provides a benchmark for developing and systematically evaluating MLFFs on soft matter systems containing experimentally measured densities and glass transition temperatures.
        PolyData (e--j) is a collection of training datasets for polymeric systems, consisting of three subsets, PolyPack, PolyDiss, and PolyCrop.
        \textbf{a}, Number of polymers containing at least one atom of the respective element.
        \textbf{b}, Distribution of molecular weights of repeating units.
        \textbf{c}, Distribution of experimental (volumetric mass) densities under standard conditions.
        \textbf{d}, Distribution of experimental glass transition temperatures \tg.
        \textbf{e}, Example structure from PolyPack, which contains structures with multiple polymer chains packed in periodic boxes at different densities.
        \textbf{f}, Distribution of the norm of the nuclear forces in PolyPack.
        \textbf{g}, Example structure from PolyDiss shown with one periodic image.
        The ``offset'' is the minimal interatomic distance across the periodic boundary.
        PolyDiss structures contain a single polymer chain in a periodic box of increasing size, highlighting the inter-chain interactions of polymers.
        \textbf{h}, Dissociation curve of a single polymer chain in a periodic box.
        \textbf{i}, Example structure from PolyCrop, which contains fragments of polymer chains in vacuum.
        \textbf{j}, Distribution of the longest dimension of the PolyCrop clusters.
        \textbf{k}, \textbf{l}, Repeating units of a selection of polymers from the PolyArena benchmark that are studied in more detail.
        Polymers that were part of the training data are shown in \textbf{k}, while those that were not are shown in \textbf{l}.
        End groups are chosen for structural stability, not synthetic feasibility.
    }
    \label{fig:data}
\end{figure}

To accelerate the adoption of MLFFs in polymer science, relevant and robust benchmarks are needed to evaluate their performance.
At the same time, MLFF development must be guided by experimental validation.
Current MLFF benchmarks primarily focus on reproducing quantum-chemical reference data, such as energies and forces, for small molecules or crystalline materials\supercite{Ramakrishnan2014,Chanussot2021OC20,Tran2023OC22,Levine2025OMol25,Gharakhanyan2025OMC25,Riebesell2025Framework}.
To this end, we introduce PolyArena, a benchmark for evaluating MLFFs on experimentally measured bulk properties of polymers.
Compared to benchmarks centered around small molecules, the vast chemical diversity spanned by their monomeric building blocks and complex intra- and intermolecular interactions present unique challenges.
This complexity, together with the breadth of available experimental data and their industrial importance, makes polymers an ideal test bed for developing and validating MLFFs.

PolyArena contains experimental densities and glass transition temperatures, \tg s, under standard conditions for \num{130} polymers from the Bicerano handbook\supercite{Bicerano2002Prediction}.
These two quantities are fundamental polymer descriptors as they determine its bulk characteristics, such as mechanical properties and thermal stability, that ultimately affect processing and application design\supercite{Rubinstein2017Polymer,Odegard2021Molecular,Odegard2022Accurate,Galante2023Anisotropic,Zhang2025London}.
The prediction of these properties through atomistic simulation requires an accurate description of the underlying intra- and intermolecular interactions\supercite{Rubinstein2017Polymer,Odegard2021Molecular,Odegard2022Accurate,Galante2023Anisotropic,Zhang2025London}.
In particular, the accurate simulation of the glass transition is a long-standing challenge in polymer modeling and an active field of research, as it requires capturing a complex interplay of local and non-local interactions across multiple length and time scales\supercite{Berthier2023Modern}.
The glass transition is commonly treated as a second-order thermodynamic transition and is identified as a change in slope of a thermophysical observable, such as the density, as a function of temperature.
During this heating-induced transition, amorphous polymers transform from a hard, brittle glassy state to a rubbery or viscous state (for more details, see Section~\ref{sec:tg_sim})\supercite{Rubinstein2017Polymer}.

Importantly, PolyArena spans polymers containing main-group elements from the first three periods, specifically H, C, N, O, F, Si, S, and Cl (Fig.~\ref{fig:data}a).
A variety of polymer families are represented, among them polyolefins, polyesters, polyethers, polyacrylates, polycarbonates, polyimides, polystyrenes, siloxanes, and perfluorinated polymers.
Decorating functional groups include alkyl chains, nitriles, carboxylic acid derivatives, and halogens (Fig.~\ref{fig:data}k, l).
Initiating and terminating groups are assigned to each polymer considering only structural compatibility, not synthetic feasibility, to reflect the dominance of the polymer body in experimental measurements.
The molecular weight of the repeating units covers a broad interval from \qtyrange{28}{593}{\gpmol}, encompassing both the simplest polyolefins and complex AB-alternating copolymers 
(Fig.~\ref{fig:data}b).
Experimental densities under standard conditions range from \qtyrange{0.8}{2.0}{\gpcc} (Fig.~\ref{fig:data}c) and
glass transition temperatures span \qtyrange{152}{672}{\K} (Fig.~\ref{fig:data}d).
Together, these distributions define a chemically and thermophysically diverse regime for benchmarking MLFFs on experimentally measurable polymer bulk properties.

To facilitate the application of MLFFs to polymers, we also introduce an accompanying collection of training datasets, called PolyData, which contains atomistic structures of polymers from PolyArena labeled with quantum-chemical methods.
Existing MLFF datasets primarily focus on small molecules or crystalline materials and are thus not necessarily well-suited for capturing the complex interplay of interactions governing polymer behavior.
PolyData is specifically designed to close this gap.
It is composed of three subsets, PolyPack, PolyDiss, and PolyCrop (Fig.~\ref{fig:data}e--j).
PolyPack and PolyDiss contain atomistic polymeric structures with periodic boundary conditions.
PolyPack contains multiple structurally-perturbed polymer chains packed at various densities.
In contrast, PolyDiss consists of single polymer chains in unit cells of varying sizes.
The two datasets are complementary, as they target different types of interactions.
While PolyPack primarily probes strong intramolecular (or intra-chain) interactions, PolyDiss focuses on weaker intermolecular (or inter-chain) ones.
PolyCrop contains fragments of polymer chains in vacuum.
For details on the generation of these datasets and the quantum-chemical methods employed for labeling, we refer the reader to Section~\ref{sec:methods}.

\subsection{Vivace -- A Fast, Scalable, and Accurate MLFF}
\label{sec:model}

Vivace is a local SE(3)-equivariant graph neural network (GNN) engineered for the speed and accuracy required for large-scale atomistic polymer simulations.
Its local architecture is based on that of Allegro\supercite{Musaelian2023Learning}, enabling efficient multi-graphics-processing-unit (GPU) calculations.
While it draws inspiration from a variety of other successful models\supercite{Unke2019PhysNet,Batzner2022Equivariant,M3GNet,Gasteiger2021GemNet,SEGNN,EGNN,GeoMFormer,shi2022benchmarking,Equiformer,EquiformerV2,simeon2023tensornet,Musaelian2023Learning},
we propose two key innovations to enhance computational efficiency while maintaining accuracy.
First, Vivace employs computationally efficient SE(3)-equivariant operations.
While crucial for accuracy, standard equivariant operations introduce a significant computational bottleneck, often involving expensive tensor products\supercite{Batzner2022Equivariant,Batatia2025Design,Batatia2022MACE}.
We find that learnable parameters within these operations are not essential for achieving high accuracy.
Instead, Vivace captures crucial three-body interactions using a lightweight tensor product and an efficient inner-product operation.
This approach is mathematically related to more complex triangular attention mechanisms\supercite{abramson2024accurate,M3GNet} but is substantially more computationally efficient.
Second, Vivace employs a multi-cutoff strategy to balance accuracy and efficiency.
Efficient, invariant operations handle weaker mid-range interactions up to \qty{6.5}{\angstrom}.
In contrast, computationally expensive equivariant operations are reserved for short-range interactions below \qty{3.8}{\angstrom}.
This strategy allows Vivace to achieve the accuracy of a large cutoff with the efficiency of a smaller one.
More model details can be found in Section~\ref{sec:model_details}.

Vivace was trained exclusively on computational data, with no experimental input.
Specifically, Vivace was first pre-trained on a subset of OMol25\supercite{Levine2025OMol25} and PolyCrop, which contain only non-periodic structures.
For subsequent fine-tuning, we introduced two datasets with periodic structures: PolyPack and PolyDiss.
To retain knowledge from pre-training, Vivace was then fine-tuned on a combination of all datasets, with a higher weight assigned to the periodic configurations (see Section~\ref{sec:hyperopt} for details)\supercite{Gardner2025Understanding}.
The importance of fine-tuning on these periodic datasets is investigated in Section~\ref{sec:interactions}.

For a more detailed analysis, we selected a diverse set of 10 polymers (Fig.~\ref{fig:data}k, l).
These polymers were divided into two groups of five.
One group, designated as ``seen'', was included in the training data.
The other group, designated as ``unseen'', was held out.
Vivace was trained on computational data for all polymers in PolyArena except for those in the unseen group.
This setup allows us to evaluate Vivace's ability to generalize to new polymers, a critical capability for a general-purpose polymer MLFF.

\subsection{From Ab Initio Data to Bulk Properties with MLFFs}

To compute the densities of polymers in PolyArena, we perform MD simulations with atomistic model systems and analyze the resulting trajectories.
We run these simulations with different FFs and compare the computed densities to experimental measurements.
Concretely, we compare Vivace to two widely-used classical FFs, the Polymer Consistent Force Field (PCFF)\supercite{Sun1994PCFF} and the refined Optimized Potentials for Liquid Simulations (OPLS) OPLS3e\supercite{Roos2019OPLS3e}, and two state-of-the-art MLFFs, MACE-OFF\supercite{Kovacs2025MACEOFF} and UMA\supercite{Wood2025UMA}.
For more details on the simulation protocol and models, we refer the reader to Section~\ref{sec:methods}.
For each model, we report the mean absolute error (MAE) for all successfully simulated polymers.
A separate MAE is also calculated for the 10 selected polymers.
For Vivace, we further distinguish between MAEs for polymers seen and unseen during training.

\begin{figure}[htb]
    \centering
    \includegraphics[width=\textwidth]{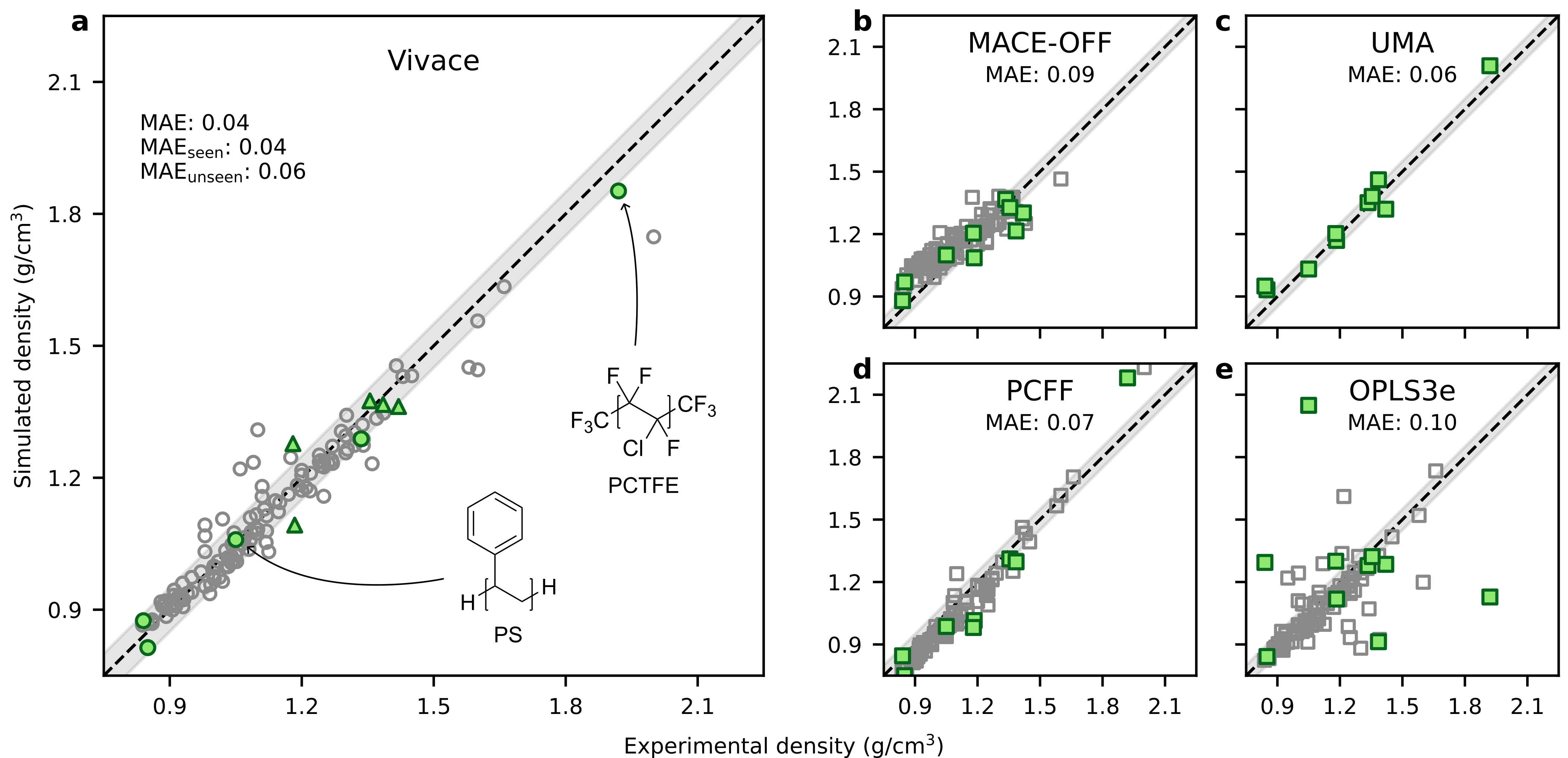}
    \caption{
        With Vivace, trained solely on \textit{ab initio} data, one can accurately predict experimental densities for a wide range of polymers.
        In each panel, we plot calculated vs experimental densities of polymers from the PolyArena benchmark under standard conditions for a different FF.
        Each marker corresponds to an MD simulation from which the density was calculated.
        A selected list of 10 polymers are indicated by green markers.
        Due to its slow simulation speed, UMA was only applied to these polymers.
        Not all polymers were simulated with the remaining models due to simulations failing (see main text for details).
        In \textbf{a}, circles ($\scalebox{0.8}{$\bigcirc$}$) and triangles ($\triangle$) indicate polymers seen and unseen during Vivace's training, respectively.
        The polymers PS (seen) and PCTFE (unseen) are annotated.
        In panels \textbf{b}, \textbf{c}, \textbf{d} and \textbf{e} we do not make this distinction and indicate all polymers with the same square marker ($\square$).
        MAEs (in \unit{\gpcc}) are reported in each panel.
        The dashed lines indicate perfect agreement, and the shaded areas show a deviation of \qty{\pm 0.05}{\gpcc}.
    }
    \label{fig:density}
\end{figure}

Fig.~\ref{fig:density} compares computed densities to experimental densities of polymers for the different models.
Not all polymers could be simulated with all models, as some simulations failed due to numerical instabilities or model incompatibilities.
For example, as MACE-OFF was not trained on silicon, it could not be applied to silicon-containing polymers.
Further, UMA was only applied to 10 selected polymers due to its slow simulation speed.
It can be seen that Vivace accurately predicts experimental densities for a wide range of polymers.
For polymers seen during training, Vivace achieves an MAE of \qty{0.04}{\gpcc} while for the five unseen polymers, its MAE increases to \qty{0.06}{\gpcc}.
Nonetheless, it indicates that Vivace can generalize to unseen polymers, which is a key requirement for a general-purpose polymer MLFF.

Overall, Vivace and UMA perform best, with an MAE of \qty{0.04}{\gpcc} and \qty{0.06}{\gpcc}, respectively, outperforming the two classical FFs, PCFF and OPLS3e, which achieve MAEs of \qty{0.07}{\gpcc} and \qty{0.10}{\gpcc}, respectively.
This is a surprising result given that UMA was not trained on any polymer data.
MACE-OFF, despite not having been trained on any polymer data, achieves an MAE of \qty{0.09}{\gpcc} on this out-of-distribution task, underscoring the transferability of this MLFF.
In conclusion, it can be seen that MLFFs trained solely on \textit{ab initio} data can accurately predict experimental densities for a wide range of polymers; experimental data do not appear necessary.
This is in contrast to the classical FFs considered in this study, the parametrization of which involves experimental data.

In Table~\ref{tab:density_mae}, \ref{tab:speed_table} and \ref{tab:speed_per_atom}, we provide a comparison between the different models in terms of density MAEs and simulation times.
It can be seen that Vivace and MACE-OFF are significantly faster than UMA; however, it should be noted that the simulation setups were not identical, so a direct comparison is difficult (see Section~\ref{sec:comp_eff} for details).
The classical FFs are faster than all MLFFs by at least an order of magnitude.

\begin{table}[H]
    \caption{%
        Mean Absolute Errors (MAEs) of computed densities with respect to experimental values over 10 selected polymers and
        MD simulation speeds for different models.
        Simulation speeds are reported with a \qty{0.5}{\fs} time step and were measured with a polystyrene system containing around \num{8000} atoms 
        on a single NVIDIA A100 GPU with \qty{80}{\giga\byte} of memory.
        See Section~\ref{sec:comp_eff} for details on the speed measurements.
        \label{tab:density_mae}
    }
    \begin{threeparttable}
        \centering
        \begin{minipage}{\linewidth}
            \sisetup{detect-weight=true}
            \begin{tabular*}{0.9\linewidth}{@{}l@{\extracolsep{\fill}}S[table-format=1.2]@{\extracolsep{15mm}}S[table-format=1.2]@{\extracolsep{15mm}}S[table-format=2.2]@{}}
                \toprule
                ~ & \multicolumn{2}{c}{MAE (\unit{\gpcc})} &       \\ \cmidrule{2-3}
                Model                                & {Overall} & {10 Polymers} & {Speed (\unit{\nspday})}\\
                \midrule
                Vivace  & 0.04 & 0.05 & 0.52                        \\
                UMA\tnote{$\diamond$}  &0.06 &  0.06 & 0.03                        \\
                MACE-OFF &0.09 & 0.08\tnote{\textdagger}&  0.51                       \\
                PCFF & 0.07 & 0.12\tnote{$\ddagger$}                &           17.68                          \\
                OPLS3e  & 0.10 & 0.31               & {---\tnote{$\ast$}} \\
                \bottomrule
            \end{tabular*}
            \begin{tablenotes}
                \small
                \item[$\diamond$] The overall MAE for UMA is based on the 10 selected polymers only.
                \item[\textdagger] MD simulation of PCTFE failed with MACE-OFF.
                \item[$\ddagger$]MD simulations of PET and PODPM failed with PCFF.
                \item[$\ast$] We do not have access to OPLS3e to measure its speed; however, we expect it to be on the same order of magnitude as that of PCFF.
                MAEs are based on results from Ref.~\cite{Afzal2021Schroedinger}.
            \end{tablenotes}
        \end{minipage}
    \end{threeparttable}
\end{table}

\subsection{Observing Glass Transitions}

We next investigate Vivace's ability to capture glass transitions, which are second-order phase transitions.
The temperature at which this transition occurs is determined computationally by simulating the density of a polymer as a function of temperature and identifying the change in slope associated with the transition.
By fitting the density-temperature curve, we extract a fitted glass transition temperature, \tgfit{}.
The uncertainty of this fit is estimated by bootstrapping the density data and their associated uncertainties (see Section~\ref{sec:tg_sim} for details).

For these experiments, we increase the system size to \num{10000} atoms to reduce the variance of the \tg{} estimates (see Fig.~\ref{fig:system_size}).
This is consistent with previous studies\supercite{Afzal2021Schroedinger} and highlights the need for MLFFs that scale to large systems.
Due to the high computational cost of these simulations, we restrict this study to the 10 polymers shown in Fig.~\ref{fig:data}k-l.
Furthermore, we exclude UMA from this experiment due to its slow simulation speed and large memory requirements.
We therefore compare Vivace only to MACE-OFF and the two classical FFs, PCFF and OPLS3e.

\begin{figure}[H]
    \centering
    \includegraphics[width=\textwidth]{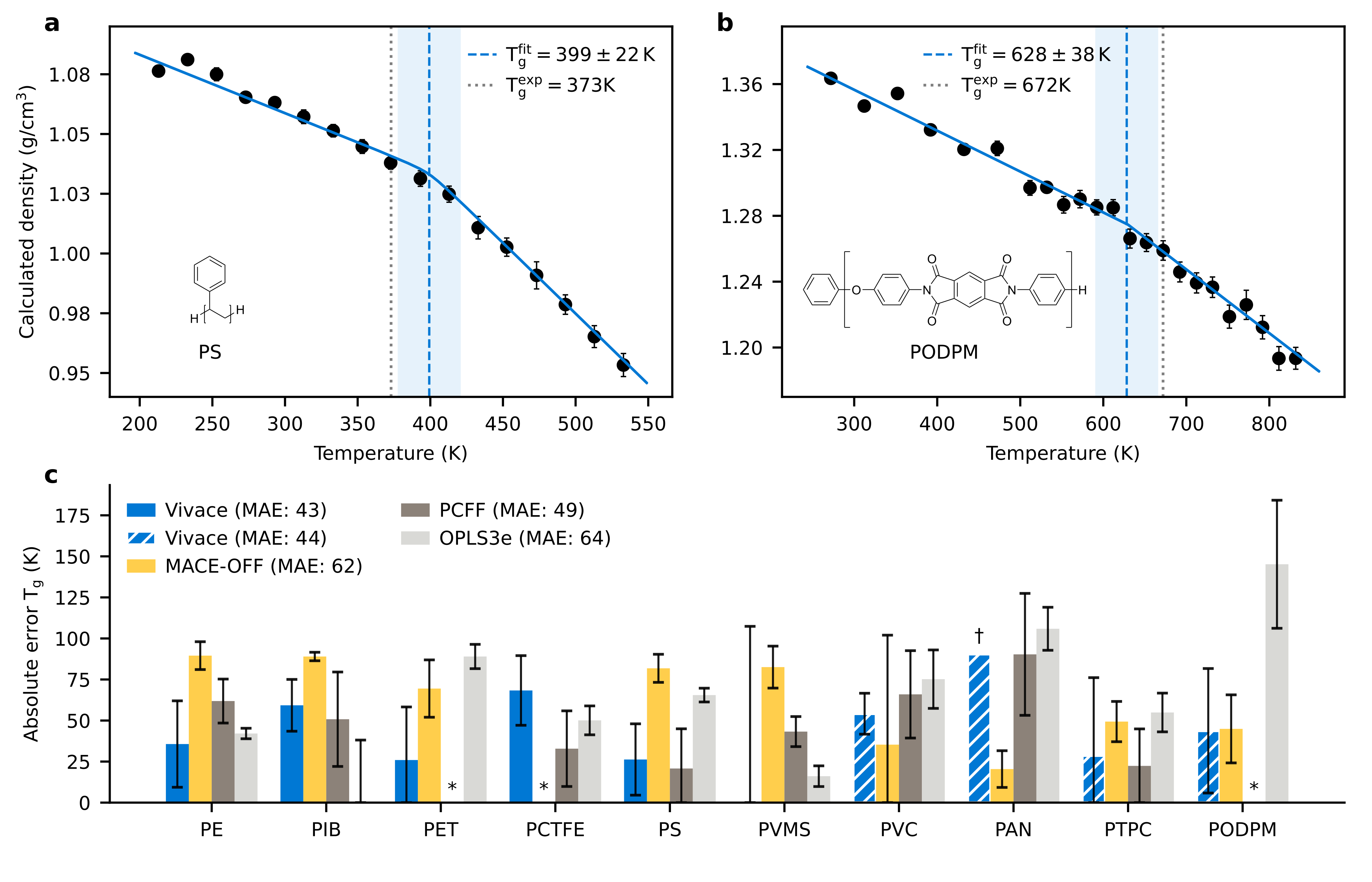}
    \caption{%
        Calculations using different FFs of the glass transition temperatures of 10 polymers selected from the PolyArena benchmark.
        \textbf{a}, \textbf{b}, Density vs temperature curves obtained for two selected polymers using Vivace.
        Each point represents the average of three density simulations, with weights inversely proportional to their uncertainty.
        Error bars indicate the combined uncertainty.
        Examples of fits used to derive \tgfit{} are shown in blue.
        \tgfit{} and the experimental \tg{} are indicated by dashed and dotted vertical lines, respectively.
        The shaded area indicates the uncertainty of the fit, derived by bootstrapping.
        More density vs temperature curves are shown in Figs.~\ref{fig:tg_mlff}, \ref{fig:tg_mace}, and \ref{fig:tg_pcff}.
        \textbf{c}, Absolute errors (and corresponding uncertainty) of calculated vs experimental \tg's for different FFs.
        For Vivace we indicate polymers that were not seen during training by a hatched pattern.
        In the case of PAN, the large uncertainty range of \qty{211}{\K} is not plotted ($\dagger$).
        Asterisks ($\ast$) indicate failed MD simulations so that no \tg{} could be determined.
        MAEs (in \unit{\K}) are reported across all polymers for which \tg{} could be determined.
    }
    \label{fig:tg}
\end{figure}

The density-temperature curves generated with Vivace generally exhibit the two linear regimes characteristic of a second-order phase transition (see, for instance, Fig.~\ref{fig:tg}a).
To our knowledge, this is the first time an MLFF has been shown to capture second-order phase transitions in polymers.
Fig.~\ref{fig:tg}b highlights the difficulty of this task, where the change in slope is less distinct, leading to greater prediction uncertainty.
The remaining density-temperature curves are provided in Fig.~\ref{fig:tg_mlff}.

From these curves, we extract \tgfit{} and compare it to the experimental \tg{} (Fig.~\ref{fig:tg}c).
For instance, Vivace estimates the \tg{} of polystyrene to be \qty{399 \pm 22}{K}, which is close to the experimental value of \qty{373}{K}.
Overall, Vivace achieves an MAE of \qty{43}{K} across all 10 polymers, with a similar performance on seen and unseen polymers (MAEs of \qty{43}{K} and \qty{44}{K}, respectively); a promising result given the task's complexity.
This accuracy is comparable to that of PCFF which has an MAE of \qty{49}{K}, with MACE-OFF and OPLS3e performing slightly worse (MAEs of \qty{62}{K} and \qty{64}{K}, respectively).
Some simulations with PCFF failed due to instabilities, preventing \tg{} determination and highlighting the limited transferability of classical FFs.
Further, we found that Vivace and MACE-OFF tend to overestimate \tg{} values, which is consistent with previous findings made for classical FFs (see Fig.~\ref{fig:tg_methods})\supercite{Gudla2024How,Afzal2021Schroedinger}.
Finally, we found that an accurate \tg{} prediction does not necessarily correlate with an accurate density prediction at room temperature, and vice versa (see Figs.~\ref{fig:tg_mace} and \ref{fig:tg_pcff}).

\section{Discussion}

\subsection{Modeling Inter-Molecular Interactions}
\label{sec:interactions}

As described in Section~\ref{sec:model}, Vivace was pre-trained on a subset of OMol25\supercite{Levine2025OMol25} and PolyCrop, both of which contain only non-periodic structures.
For fine-tuning, we introduced two datasets with periodic structures: PolyPack and PolyDiss.
We found that MD simulations with the pre-trained Vivace model yield densities substantially lower than their corresponding experimental values, resulting in a large density MAE of \qty{0.60}{\gpcc} (Fig.~\ref{fig:density_pretrained}).
One might hypothesize that the poor performance stems from the model not being trained on periodic data, leading to inaccurate stress tensor predictions.
However, MACE-OFF, which was also not trained on periodic data, achieves a much smaller density MAE of \qty{0.10}{\gpcc} than the pre-trained Vivace model.
The results shown in Table~\ref{tab:polydata} do not support this hypothesis either.
The table summarizes the performance on the PolyPack and PolyDiss test sets for the pre-trained and fine-tuned Vivace models, as well as for MACE-OFF and UMA.
The fine-tuned Vivace model's stress (and force) MAE on PolyPack is comparable to that of the pre-trained model and MACE-OFF.
This observation highlights how standard computational metrics, such as force and stress errors, are not always good predictors of a model's ability to run stable MD simulations\supercite{Mannan2025Evaluating,Han2025Benchmarking}.

\begin{figure}[htb]
    \centering
    \includegraphics[width=\textwidth]{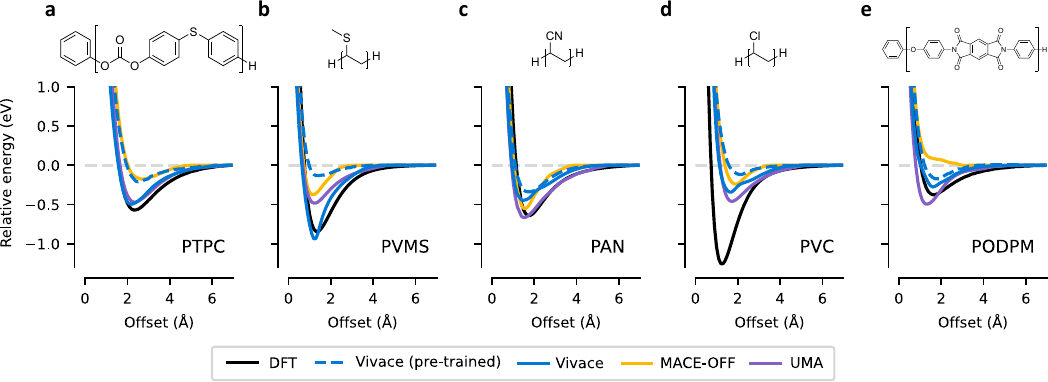}
    \caption{
        Fine-tuning Vivace improves its ability to describe inter-chain interactions.
        The figure shows selected dissociation curves from the PolyDiss test set, computed with four MLFFs and the density functional theory (DFT) reference method (black).
        Each panel plots the relative energy (in \unit{\ev}) as a function of the distance (offset) between polymer chains (in \unit{\angstrom})
        (see Section~\ref{sec:poly_data} for details).
        To account for differences in absolute energies between models, the curves are shifted vertically so that the energy at the largest separation is zero.
    }
    \label{fig:dissociation}
\end{figure}

The dissociation curves for the PolyDiss test set (Fig.~\ref{fig:dissociation}) suggest an alternative explanation.
The pre-trained Vivace model generally underestimates inter-chain dissociation energies, whereas the fine-tuned model more closely follows the quantum-chemical reference.
We therefore hypothesize that the pre-trained model's poor MD performance stems from its inadequate description of inter-chain interactions.
This hypothesis is further supported by the following observation: MACE-OFF, which predicts repulsive interactions for PODPM (Fig.~\ref{fig:dissociation}e), yields a density that is far below the experimental value (Fig.~\ref{fig:tg_mace}b).
It seems that a qualitatively accurate description of intermolecular interactions is a necessary but not sufficient condition for reliable simulations.

\begin{table}[htb]
    \centering
    \caption{
        Standard computational metrics do not always predict performance in MD simulations.
        This table shows MAEs for nuclear forces (in \unit{\mevpa}) and stress components (in \unit{\mevpac}) on the PolyPack and PolyDiss test sets.
        Density MAEs (in \unit{\gpcc}) from MD simulations are also included for comparison.
        The PolyPack and PolyDiss datasets primarily probe intra- and inter-molecular interactions, respectively.
        The Vivace model was pre-trained on a subset of the OMol25 dataset\supercite{Levine2025OMol25} and non-periodic polymer clusters.
        It was then fine-tuned by adding the periodic PolyPack and PolyDiss datasets (see Section~\ref{sec:hyperopt}).
        Results are compared to two other MLFFs, MACE-OFF\supercite{Kovacs2025MACEOFF} and UMA\supercite{Wood2025UMA}.
    }
    \sisetup{table-format = 3.1, table-alignment-mode = format}
    \setlength{\tabcolsep}{0pt}
    \begin{tabular*}{\linewidth}{@{}@{\extracolsep{\fill}}lccS[table-format=3.0]S[table-format=2.1]c@{}}
        \toprule
        & \multicolumn{2}{c}{PolyPack} & \multicolumn{2}{c}{PolyDiss} & \multicolumn{1}{c}{PolyArena}                          \\
        \cmidrule{2-3} \cmidrule{4-5} \cmidrule{6-6}
        {Model}                & {Force MAE}                   & {Stress MAE}                   & {Force MAE} & {Stress MAE} & {Density MAE}\\
        \midrule
        Vivace               & 183                         & 1.9                          & 97        & 1.9       & 0.04 \\
        Vivace (pre-trained) & 179                         & 1.7                          & 123       & 2.4       & 0.60 \\
        MACE-OFF           & 188                         & 3.1                          & 242       & 19.0      &  0.10 \\
        UMA                  & 110                         & 1.0                          & 108       & 1.8     &  0.05 \\
        \bottomrule
    \end{tabular*}
    \label{tab:polydata}
\end{table}

\subsection{Effect of the Cutoff Radius}
\label{sec:cutoff}

MLFFs operate on a three-dimensional graph of atoms, where edges connect atoms within a specified cutoff radius.
A larger cutoff radius allows the model to capture interactions over greater distances.
However, the number of edges scales cubically with the cutoff radius, making large cutoff radii prohibitively expensive.
The cutoff radius is distinct from the receptive field, which is the maximum distance over which information can propagate through multiple message-passing iterations.
It has been shown that a sufficiently large graph cutoff is necessary for an accurate force field, and that a large receptive field does not compensate for a small cutoff\supercite{kovacs2023evaluation,chmiela2023accurate}.
Large receptive fields also lead to poor parallelization across multiple devices, which becomes relevant to large-scale simulations involving tens of thousands of atoms.
Consequently, choosing the cutoff radius involves a trade-off between accuracy and computational efficiency.

\begin{figure}[htb]
    \centering
    \includegraphics{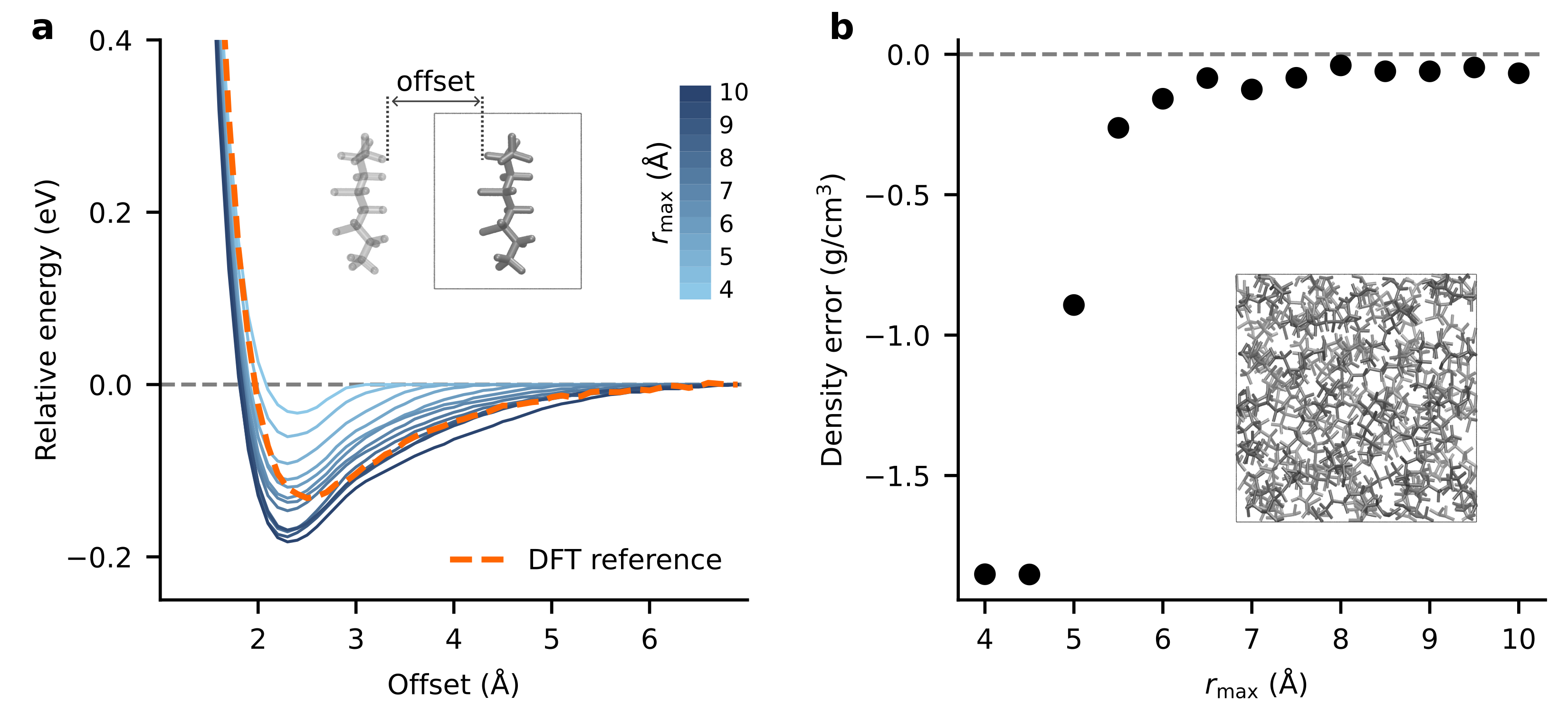}
    \caption{%
        A sufficiently large cutoff radius is essential for accurate polymer simulations with MLFFs.
        \textbf{a}, Dissociation curves for polychlorotrifluoroethylene (PCTFE) from the PolyDiss dataset, computed with Vivace models having different cutoff radii ($\rmax$).
        The DFT reference is shown as a dashed orange line.
        The plot shows the relative energy (in \unit{\ev}) as a function of the distance (offset) between polymer chains (in \unit{\angstrom}).
        To account for differences in absolute energies between models, each curve is shifted vertically so that its energy at the largest separation is zero.
        The inset shows a polymer chain and one of its periodic images.
        \textbf{b}, Error in PCTFE densities derived from MD simulations using Vivace models with varying $\rmax$.
        A snapshot of a simulation system is shown in the inset.
        For cutoff radii below \qty{5.5}{\angstrom}, the polymer chains repel each other, causing the system's density to collapse.
    }
    \label{fig:cutoff}
\end{figure}

We investigated the effect of the cutoff radius, especially as Vivace's receptive field is equal to its cutoff radius by design (see Section~\ref{sec:model_details}).
Fig.~\ref{fig:cutoff}a shows that a cutoff radius below \qty{6}{\angstrom} fails to capture the binding energy between PCTFE polymer chains, predicting a too weak attraction that vanishes prematurely.
This observation on a molecular level is consistent with the density predictions from MD simulations.
From Fig.~\ref{fig:cutoff}b, we see that MD simulations using Vivace models with a cutoff radii below \qty{5.5}{\angstrom} results in density predictions that are far below the experimental values.
Through visual inspection of the simulation trajectories, we found that the polymer chains unfold and repel each other, causing the system's density to collapse to near-zero values.
A cutoff of approximately \qty{6.5}{\angstrom} seems to be required for the simulated density to approach the experimental value.
This finding is consistent with the observation that MD simulations of PCTFE with MACE-OFF, which has a cutoff of \qty{4.5}{\angstrom}, suffer from a similar density collapse (Fig.~\ref{fig:tg_mace}b).
By contrast, UMA has a cutoff of \qty{6}{\angstrom} and does not exhibit this issue.
We found similar trends for other polymers, with PCTFE being particularly sensitive to this parameter, but expect this cutoff threshold to vary between polymers.

Vivace's architecture aims to strike a balance between accuracy and efficiency by introducing a multi-cutoff strategy.
Longer-range interactions within \qty{6.5}{\angstrom}, which are typically weak and can be approximated by simple two-body functions (e.g., van der Waals interactions), are modeled by efficient, invariant operations.
Then, computationally expensive operations operating on higher-order geometric tensors are only applied to model short-range interactions (\qty{< 3.8}{\angstrom}).
This way Vivace can achieve the accuracy enabled by a large cutoff while maintaining the efficiency of a smaller one.

\section{Conclusions and Outlook}

We demonstrated that an MLFF, trained exclusively on \textit{ab initio} data, accurately predicts experimental bulk properties for a wide range of polymers.
Our approach outperforms traditional classical FFs, which rely on extensive, experiment-based parameterization.
Furthermore, we showed that MLFFs can capture second-order phase transitions in soft matter, enabling the prediction of properties like the glass transition temperature.

Several technical advancements made these findings possible.
First, we introduced PolyArena, the first large-scale benchmark for MLFFs based on experimentally measured bulk properties of polymers.
This benchmark is designed to complement existing computational ones and encourage the community to pursue more experimental validation.
Second, we provide PolyData, a collection of datasets for training and fine-tuning MLFFs for polymer science applications.
With broad chemical coverage and highly curated quantum-chemical labels, PolyData enables models to better generalize across a wide range of systems and conditions.
Finally, we developed the Vivace architecture, a new, fast SE(3)-equivariant GNN optimized for efficient, large-scale simulations.

To further advance MLFFs for polymers, several challenges remain.
Non-covalent long-range interactions, such as electrostatics, need to be better represented, as they are crucial in many polymers\supercite{Hribar-Lee2014Polyelectrolytes,Poelking2015Impact}.
The MLFFs used in this work do not explicitly model these interactions, but recent progress in this area is promising\supercite{Ko2021Fourthgeneration,Zinovjev2023Electrostatic,Gubler2024Accelerating,Loche2025Fast,Cheng2025Latent}.
Additionally, predicting polymer properties with MLFFs remains computationally intensive.
For classical FFs to be replaced by data-driven alternatives in non-reactive scenarios, improvements in computational efficiency are needed.

This work highlights the transformative potential of MLFFs for polymer simulations and lays the groundwork for future research.
Upcoming studies should expand this approach to other bulk properties, such as elastic moduli and thermal expansion coefficients.
Most importantly, future iterations of polymer-focused MLFFs will transform applications that have been challenging for traditional methods, particularly in studying reactive processes like polymer degradation, chemical recycling, and dynamic cross-linking.

\section{Methods}
\label{sec:methods}

\subsection{Details on Experimental Data}
\label{sec:poly_arena}

The experimental measurements included in PolyArena have some margin of error associated with their collection.
First, the density and glass transition temperature of a given polymer sample can be affected by the conditions under which it was prepared and processed\supercite{Bicerano2002Prediction}.
To limit the effects of sample preparation, the experimental measurements aggregated by Bicerano and employed in PolyArena represent amorphous, atactic specimens produced under standard synthesis and post-processing conditions to produce bulk form factors, avoiding those fabricated via aerosol synthesis, hot pressing, or related techniques.
Second, measured \tg{} values are dependent on the heating rate, making them particularly sensitive to the measurement technique.
This was observed to contribute \qty{2}{K} measurement error and \qty{7}{K} systematic error in the determination of the \tg{} for polystyrene\supercite{Rieger1996Glass}.

The polymers in PolyArena are extracted from the Bicerano handbook\supercite{Bicerano2002Prediction}.
We first excluded polymers with structural ambiguities or those that failed parameterization with PCFF\supercite{Sun1994PCFF} using Enhanced Monte Carlo\supercite{intVeld2003EMC}.
We also excluded polymers that were computationally challenging for our quantum-chemical methods.
Additionally, bromine-containing polymers were removed because they were incompatible with one of the quantum-chemical basis sets.
These filters reduced the initial set of polymers by approximately 10\%.
While the handbook provides densities for all polymers, \tg{} values are available for only 87 of them.

\subsection{Classical Force Field Baselines}
\label{sec:classical_ff}

We consider two classical FFs, PCFF\supercite{Sun1994PCFF} and OPLS3e\supercite{Roos2019OPLS3e} as baselines.
All densities computed with PCFF\supercite{Sun1994PCFF} use LAMMPS\supercite{Thompson2022LAMMPS} as an MD engine, with parameters provided by the Enhanced Monte Carlo (EMC) program\supercite{intVeld2003EMC}.
We do not run simulations with OPLS3e but use the densities and glass transition temperatures, together with associated uncertainties, reported in the supplementary materials of the study by \citeauthor{Afzal2021Schroedinger}\supercite{Afzal2021Schroedinger}
We note that the computational setup and analysis protocol employed in that study are similar but not identical to ours, and therefore, we exercise caution when making direct comparisons.

\subsection{Machine-learning Force Field Baselines}
\label{sec:mlff}

In this study, we restrict ourselves to two widely-used MLFF families, MACE-OFF23\supercite{Kovacs2025MACEOFF} and UMA\supercite{Wood2025UMA}, using the authors' released checkpoints without additional retraining or hyperparameter adaptation.
MACE-OFF23\supercite{Kovacs2025MACEOFF} is a family of purely short-range MLFFs for neutral, closed-shell organic systems, built on the higher-order equivariant MACE\supercite{Batatia2022MACE} architecture and is capable of accurately predicting a wide variety of gas- and condensed-phase properties of molecular systems.
MACE-OFF23 models are trained on energies and forces using an augmented version of SPICE\supercite{Eastman2023SPICE} v1 plus fragments and water clusters from QMugs\supercite{Isert2022QMugs} covering the following ten elements: H, C, N, O, F, P, S, Cl, Br, and I.
There are three size variants (\texttt{s}, \texttt{m}, and \texttt{l}) and in this study, we use the fastest variant, \texttt{s}, with 64-bit floating-point precision, unless stated otherwise.
For brevity, we refer to this model as MACE-OFF.
Models of this family have two message-passing layers, and thus, an effective receptive field of $2 \times \rmax$, where $\rmax$ is the maximum distance between two neighboring atoms.
For the \texttt{s} variant, $\rmax = \qty{4.5}{\angstrom}$. 

UMA\supercite{Wood2025UMA} is a family of so-called universal MLFFs that aim to be applicable to chemically and structurally diverse domains such as organic molecules, molecular crystals, inorganic materials, liquids, and biomolecular fragments.
Broad coverage of chemical space (more than 80 elements) is achieved by training on energies, forces, and stresses from multiple curated first-principles datasets spanning broad element combinations and bonding motifs\supercite{Barroso-Luque2024OMat24,Chanussot2021OC20,Levine2025OMol25,Gharakhanyan2025OMC25,Sriram2025OpenDAC25}.
In this work, we use the smallest variant, \texttt{UMA-s-1p1}, together with the author-provided weights most suitable for organic systems (\texttt{task\_name=omol}).
\texttt{UMA-s-1p1} has $\rmax = \qty{6.0}{\angstrom}$ and four message-passing layers, resulting in an receptive field of \qty{24.0}{\angstrom}.

\subsection{Simulation Speed Measurements}
\label{sec:comp_eff}

MD simulations with PCFF, Vivace, and MACE-OFF were performed using LAMMPS.
The MLFFs leveraged the fast ML-IAP interface with KOKKOS developed by NVIDIA.
Since UMA does not have a LAMMPS interface, we use i-PI\supercite{Litman2024IPI} as the MD engine instead.
As i-PI, unlike LAMMPS, does not support multi-GPU simulations, UMA's simulation speed is only measured on a single GPU.
Since i-PI is not nearly as optimized as LAMMPS, the speed measurements for UMA are not directly comparable to those of Vivace and MACE-OFF.

For speed measurements, we use the protocol detailed in Section~\ref{sec:density_sim} to generate polystyrene structures at a density of 1.0 g/cm$^3$ with \num{2000}, \num{4000}, \num{8000}, and \num{16000} atoms.
The generated configurations are then used in a \qty{2.5}{\ps} micro-canonical-ensemble (NVE) simulation with a \qty{0.5}{\fs} time step.
The simulation speeds are then reported in \unit{\nspday}, representing the simulated nanoseconds per day of wall-clock time.
All speed tests are performed on a compute node with 8 NVIDIA A100(-SXM4) GPUs (\qty{80}{\giga\byte} each) connected via NVLink, ensuring all multi-GPU simulations run on the same node.

\subsection{MD Simulations for Density Calculations}
\label{sec:density_sim}

We build a periodic cell containing roughly \num{2000} atoms arranged as six chains of equal length at an initial density of \qty{0.5}{\gpcc} with EMC\supercite{intVeld2003EMC}.
Chain termini are capped with fragments resembling the repeating unit to ensure chemical stability.
The EMC-generated structure is further equilibrated using a 21-step MD simulation protocol.
The aim of this protocol is to efficiently relax amorphous polymer packings to realistic densities.
It consists of seven cycles, each with three stages: high-temperature canonical-ensemble (NVT) simulation at $T_\text{max}$, NVT simulation at the target temperature, and isothermal-isobaric-ensemble (NPT) simulation at the target temperature with pressure ramping or release.
An NVT simulation keeps the number of particles, the volume, and the temperature fixed, whereas an NPT simulation keeps the number of particles, the pressure, and the temperature fixed while letting the volume adjust so the target pressure is maintained.
Over the first three cycles the pressure is ramped up to $P_\text{max}$, and over the final four cycles it is reduced to standard pressure.
$T_\text{max}$ is set to \qty{700}{K} for all polymers to ensure it is higher than the highest \tg\supercite{Abbott2013Polymatic}.
$P_\text{max}$ must be sufficiently high to compact initially rigid configurations.
Following prior work\supercite{Larsen2011Molecular}, we use $P_\text{max} = \qty{5E4}{\bar}$ for all systems.

MD simulations run with Vivace, MACE-OFF, and PCFF use the 21-step protocol detailed above with LAMMPS as the simulation engine, where a time step of \qty{0.5}{\fs} was used.
NVT and NPT simulations used a Nos\'e-Hoover thermostat with a damping value of $\tau = \qty{50}{\fs}$.
In NPT simulations, we employed a barostat (extended Nos\'e-Hoover by \citeauthor{Martyna1999Molecular}\supercite{Martyna1999Molecular}) with a time constant of $\tau = \qty{500}{\fs}$.
We employed i-PI\supercite{Litman2024IPI} to perform MD simulations with UMA, as a LAMMPS interface was not available.
A time step of \qty{0.5}{\fs} was used.
NVT simulations used a Langevin thermostat with a friction coefficient of $\tau = \qty{50}{\fs}$.
NPT simulations employed the same Nos\'e-Hoover barostat\supercite{Martyna1999Molecular} with time constant $\tau = \qty{1000}{\fs}$ coupled to a Langevin thermostat with friction coefficient $\tau = \qty{100}{\fs}$.

\subsection{Determination of the Glass Transition Temperature Through Simulation}
\label{sec:tg_sim}

The glass transition is a second-order phase transition during which a polymer transforms from a hard, brittle glassy state into a viscous, rubbery one.
This transition is marked by a change in the slope of temperature-dependent thermophysical properties, such as density, thermal expansion coefficient, and heat capacity.
While intrinsic to amorphous polymers, this transition also occurs in the amorphous regions of semicrystalline polymers.
The presence of crystalline domains can, however, affect the observed transition temperature.

Experimentally, glass transition temperatures are measured using techniques that operate on time scales of minutes to hours.
This methodological diversity can yield different \tg{} values for the same polymer; for instance, \tg{} is known to depend on the cooling rate\supercite{Montserrat2005Effect}.
Furthermore, these experimental time scales are orders of magnitude longer than those accessible to MD simulations.
These factors present significant challenges when evaluating the accuracy of computationally derived \tg{} values.
Extracting \tg{} from simulation data also introduces fitting uncertainties that require proper quantification for reliable predictions\supercite{Suter2025Rapid}.

We determined \tg{} from density-temperature profiles using the protocol described below.
While more sophisticated methods exist for both simulation and fitting\supercite{Suter2025Rapid}, our approach was chosen because it is fully automated and requires no data filtering.
We extracted \tg{} values from density-temperature profiles generated with the protocol from Section~\ref{sec:density_sim}.
For each polymer, we simulated densities at 17 temperatures, spaced at \qty{20}{K} intervals, centered on the experimental \tg{}.
This temperature range spanned from $\tg - \qty{160}{K}$ to $\tg + \qty{160}{K}$, and was extended to include room temperature in the case of PODPM.
Simulations were not performed below \qty{100}{K}.
At each temperature, we ran three simulations with \num{10000} atoms, each starting from a different initial configuration.
This approach reduced the density variance, particularly at higher temperatures (see Fig.~\ref{fig:system_size} in the Appendix).
To derive \tg{}, we employed a bootstrap resampling method.
First, we aggregated the densities from each replica using inverse variance weighting.
The variance of the density for each replica was estimated from the fluctuations in the NPT trajectory from a production stage of at least \qty{0.5}{\ns} following the end of the 21-step density protocol.
Next, we generated \num{1000} bootstrap samples by drawing from a normal distribution centered on the aggregated mean and with the aggregated variance for each temperature point.
We then fitted a hyperbolic function to each bootstrap sample, as described by \citeauthor{Patrone2016Uncertainty}\supercite{Patrone2016Uncertainty}. This automated approach does not require user input and is considered the method of choice for deriving \tg{} from MD simulations\supercite{Afzal2021Schroedinger,Suter2025Rapid}.
Finally, we calculated the mean and standard deviation of the resulting bootstrapped \tg{} values to obtain the final estimate and its uncertainty. For robustness, we also investigated fitting a piecewise linear model using the \texttt{pwlf} library\supercite{pwlf} and observed no significant differences in the derived \tg{} values (see \ref{fig:tg_methods}).

\subsection{Details on Computational Data}
\label{sec:poly_data}

The PolyData collection of datasets consists of three subsets: PolyPack, PolyDiss, and PolyCrop.
To build PolyPack, we generate initial structures containing two polymer chains with EMC\supercite{intVeld2003EMC} for each polymer in PolyArena.
We sample different sizes (50, 100, and 250 atoms) and a broad range of initial densities spanning \qtyrange{0.1}{1.4}{\text{times}} the experimental density to promote diversity.
Each structure is subjected to a short \qty{50}{\ps} NVT simulation with LAMMPS\supercite{Thompson2022LAMMPS} at \qty{1200}{\K} using PCFF\supercite{Sun1994PCFF}.
The final frame is randomly perturbed by adding Gaussian noise with a standard deviation of \qty{0.08}{\angstrom} to all atomic coordinates.
This step introduces additional structural diversity and helps the model learn to relax highly energetic structures.
We then perform eight geometry optimization steps, storing energies, forces, and stress tensors at every step.
PolyPack contains approximately \num{230000} configurations covering all polymers in the PolyArena benchmark.

PolyDiss comprises dissociation trajectories of isolated short polymer chains to probe inter-chain interactions under periodic boundary conditions.
For each polymer in PolyArena, we built a polymer chain with three repeating units with EMC, placed it in an orthorhombic periodic cell, aligned it with the $x$-axis, and uniformly shrank the box ensuring that all inter-image atomic distances exceeded \qty{1.5}{\angstrom}.
Dissociation paths were then generated by incrementally enlarging the cell in steps of \qty{0.1}{\angstrom} along $y$, $z$, $yz$, and $xyz$ directions until the minimum inter-image separation reached \qty{7.0}{\angstrom}.
Single-point calculations were performed for every configuration, and energies, forces, and stress tensors were recorded.
PolyDiss comprises about \num{50000} configurations covering all polymers in the PolyArena benchmark.

PolyCrop contains non-periodic structures and was generated using a procedure similar to that of PolyPack.
The main difference is that large, periodic structures are cropped into non-periodic spherical clusters.
Each cluster is centered on a randomly selected atom and includes all other atoms within a given radius, often spanning segments from multiple polymer chains.
During this truncation, only C--C single bonds that cross the sphere's boundary are cleaved.
Atoms outside the sphere connected by other bond types, such as double bonds, are kept to preserve the local chemical environment.
Finally, all cleaved bonds are capped with hydrogen atoms.
We found that this cropping process can produce highly energetic configurations.
The resulting dataset contains approximately \num{800000} clusters.

\subsection{Quantum-Chemical Calculations}
\label{sec:qchem}

For the PolyPack and PolyDiss dataset, single point calculations and structure optimizations were performed with the CP2K\slash QUICK\-STEP package\supercite{VandeVondele2005Quickstep,Kuhne2020CP2K},
using mixed Gaussian and plane-wave basis sets and norm-conserving GTH pseudopotentials\supercite{Lippert1997Hybrid,Goedecker1996Separable}.
TZV2P basis functions were used for the expansion into the primary Gaussian basis set and a cutoff of \qty{1200}{\text{Ry}} was used for the auxiliary plane-waves expansion. 
The exchange-correlation energy was computed with the r$^2$SCAN functional\supercite{furness2020accurate} along with the D3 correction and with Becke-Johnson damping to account for dispersion interactions\supercite{Grimme2010Consistent,Grimme2011Effect}.

In the PolyCrop dataset, configurations were labeled with different quantum-chemical methods.
Quantum mechanical calculations were performed with the ORCA program\supercite{neese2020orca}, using the range-separated hybrid $\omega$B97X-V density functional\supercite{Mardirossian2014wB97XV} and the def2-TZVPP basis set\supercite{weigend2006accurate}.
All structures were treated as closed-shell, with a spin multiplicity of one.
If a structure was too large to be labeled with DFT within a day, it was labeled with the semi-empirical GFN2-xTB tight-binding method\supercite{bannwarth2019gfn2} instead.
Approximately \num{400000} clusters containing up to \num{234} atoms were labeled with ORCA, and another \num{400000} clusters with up to \num{1411} atoms were labeled with GFN2-xTB.

\section*{Code and Data Availability}

Release of source code and data is in preparation.

\section*{Acknowledgements}

We gratefully acknowledge the Microsoft Research AI for Science team for invaluable discussions, feedback, and support.
Especially we would like to thank (in alphabetical order):
Rianne van den Berg,
Michael Gastegger,
Paola Gori-Giorgi,
Chin-Wei Huang,
Sarah Lewis,
Frank Noé,
Ryota Tomioka,
Leo Xia,
Tian Xie,
Yu Xie,
and
Claudio Zeni.
We thank Vahideh Alizadeh for assistance with CP2K simulations and NVIDIA for technical support related to the ML-IAP interface in LAMMPS.

\printbibliography

\clearpage

\section*{Supplementary Materials}

\setcounter{section}{0}
\setcounter{table}{0}
\setcounter{figure}{0}
\setcounter{page}{0}
\renewcommand{\thesection}{\Alph{section}}
\newrefsection
\renewcommand{\thepage}{S\arabic{page}}
\renewcommand{\theequation}{S\arabic{equation}}
\renewcommand{\thefigure}{S\arabic{figure}}
\renewcommand{\thetable}{S\arabic{table}}
\renewcommand{\thealgorithm}{S\arabic{algorithm}}

\newcommand{\vincludegraphics}[2][]{%
    \parbox[c][15mm][c]{45mm}{%
        \centering
        \includegraphics[#1]{#2}%
    }%
}

\section{Selected Polymers in PolyArena}
\label{sec:selected_polymers}

\begin{table}[H]
    \caption{%
        Abbreviation, chemical structure, and name of 10 polymers selected from the PolyArena benchmark for detailed study in this work.
        The ``Seen during training'' column indicates whether the polymer was part of the training set for the Vivace model.
    }
    \label{tab:poly_ids}
    \centering
    \begin{tabular*}{\linewidth}{@{}p{15mm}@{}x{45mm}@{}l@{}r@{}}
        \toprule
        Abbreviation & Structure & Name & \makebox[15mm][r]{Seen during training} \\
        \midrule
        PE & \vincludegraphics[width=15mm]{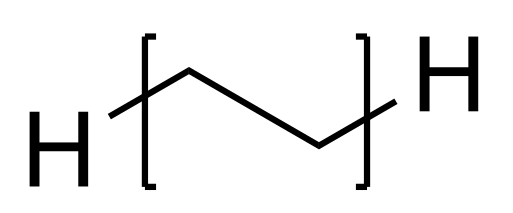} & Polyethylene & Yes \\
        PIB & \vincludegraphics[width=15mm]{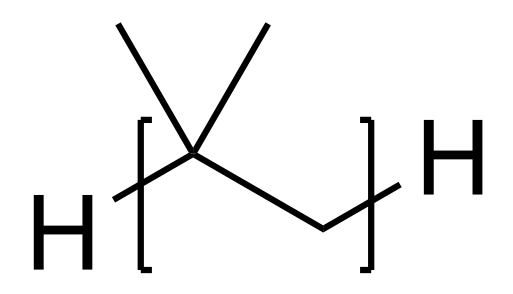} & Polyisobutylene & Yes \\
        PET & \vincludegraphics[width=35mm]{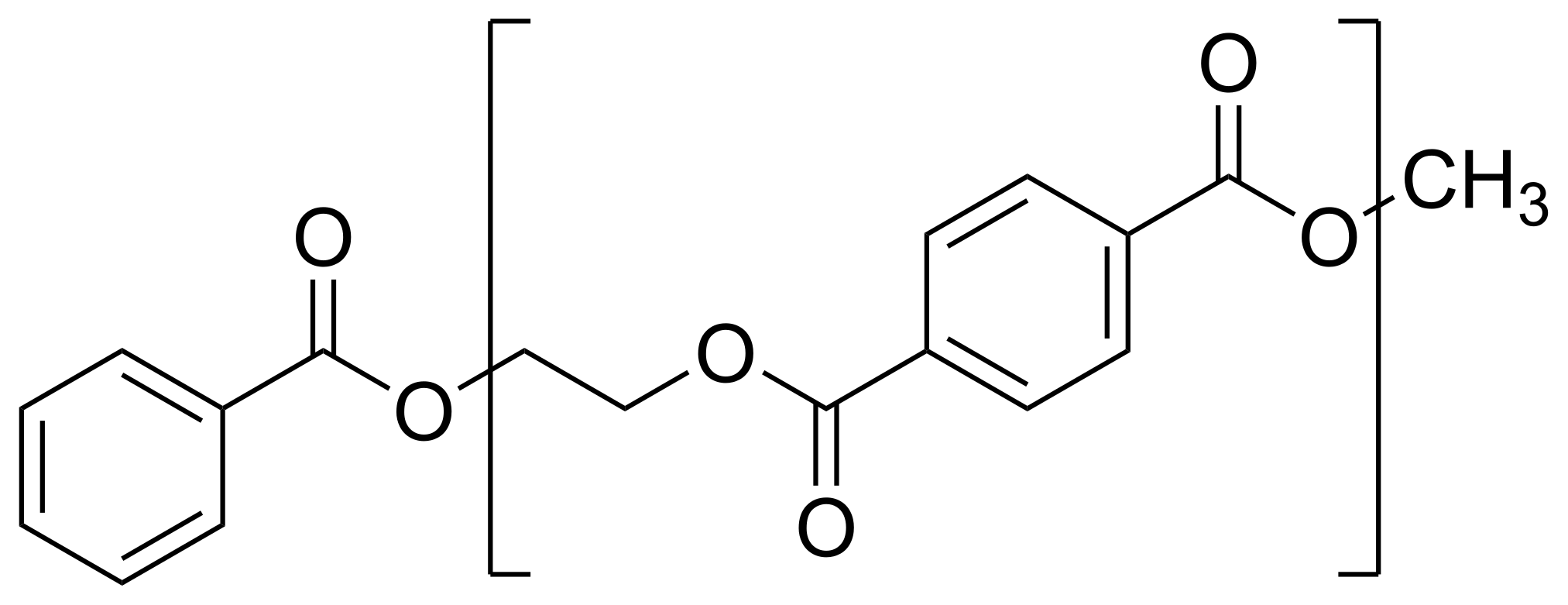} & Poly(ethylene terephthalate) & Yes \\
        PCTFE & \vincludegraphics[width=20mm]{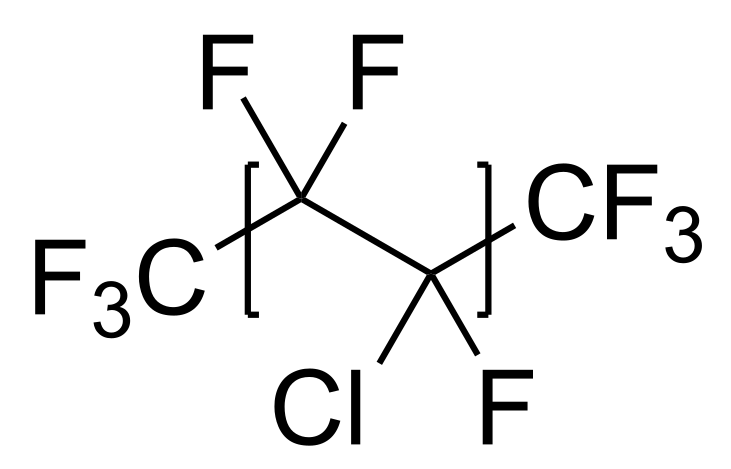} & Polychlorotrifluoroethylene & Yes \\
        PS & \vincludegraphics[width=12mm]{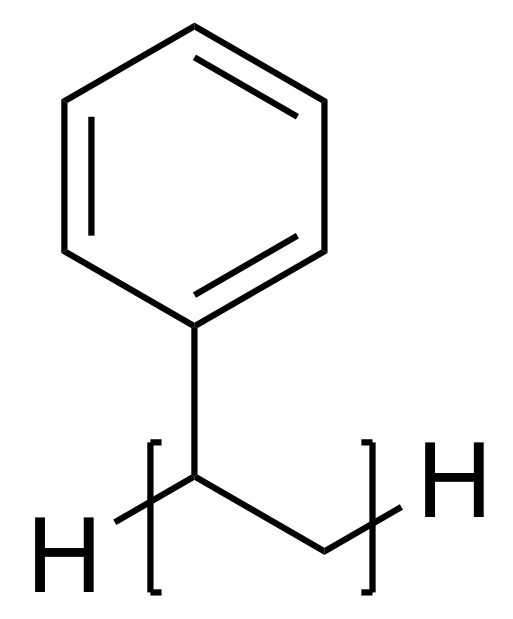} & Polystyrene & Yes \\
        PVMS & \vincludegraphics[width=15mm]{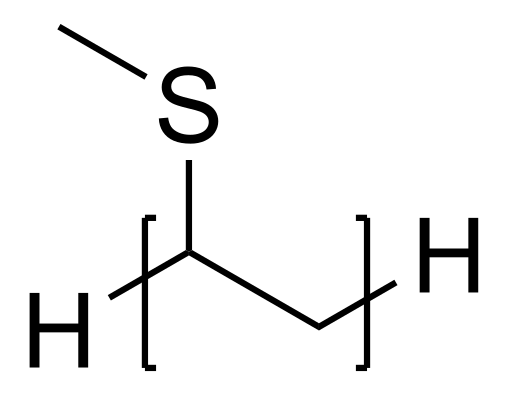} & Poly(vinyl methyl sulfide) & No \\
        PVC & \vincludegraphics[width=15mm]{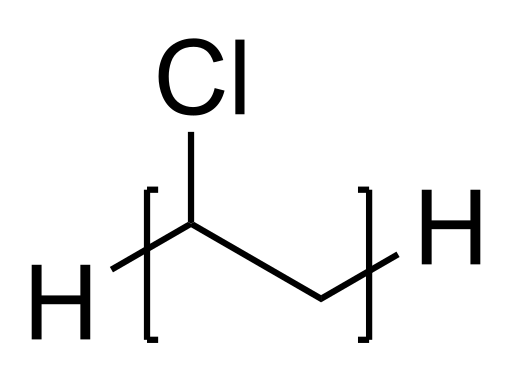} & Poly(vinyl chloride) & No \\
        PAN & \vincludegraphics[width=15mm]{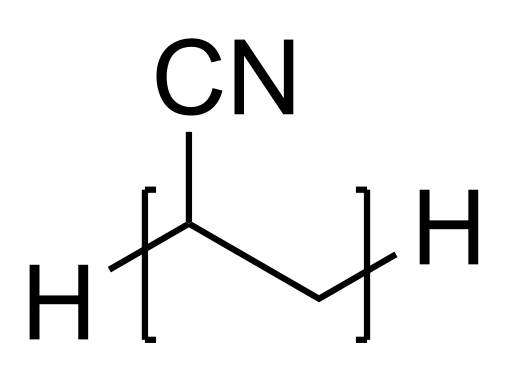} & Polyacrylonitrile & No \\
        PTPC & \vincludegraphics[width=40mm]{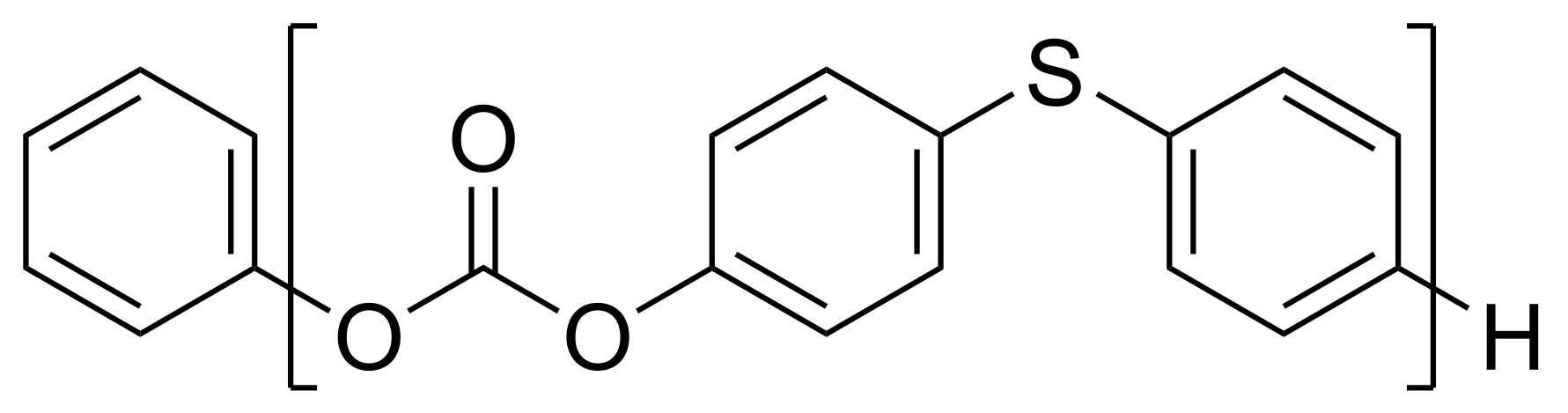} & Poly[thio bis(4-phenyl)carbonate] & No \\
        PODPM & \vincludegraphics[width=40mm]{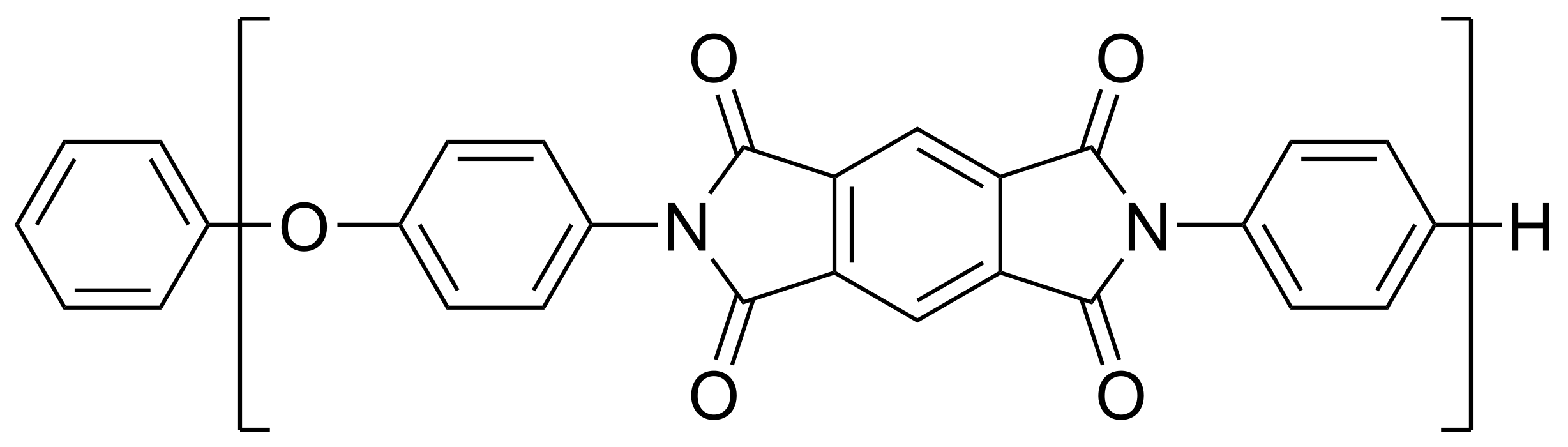} & Poly[N,N'-(p,p'-oxydiphenylene)pyromellitimide] & No \\
        \bottomrule
    \end{tabular*}
\end{table}

\section{Model Details}
\label{sec:model_details}

Since the first application of neural networks to FFs by \citeauthor{blank1995neural} in 1995,
MLFFs have grown into a powerful tool for predicting chemical reaction rates and structural evolution, which are fundamental to computational materials science, chemistry, and biology.
To date, there are many MLFF architectures \supercite{Batatia2022MACE,Batzner2022Equivariant,Musaelian2023Learning,Smith2017ANI1,DimeNet,schutt2021equivariant} and so-called foundation models.
However, the central challenge, the GAS (Generalization, Accuracy, and Speed) trilemma, remains unsolved.

Achieving a balance among these three factors is difficult, as improving two typically compromises the third.
For example, classical FFs \supercite{MMFF94,Sun1994PCFF,Roos2019OPLS3e} are fast and broadly applicable but suffer from limited accuracy.
Descriptor-based MLFFs \supercite{Behler2007Generalized,SNAP,DPMD,sGDML,MTP,ETNP}
offer improved accuracy and speed due to their relatively simple structures, yet they often lack generalization, necessitating the training of individual models for each system.
Graph neural network (GNN)-based models\supercite{Schutt2017Quantum,M3GNet,Gasteiger2021GemNet}, particularly equivariant ones\supercite{Schutt2017Quantum,Musaelian2023Learning,Equiformer,GeoMFormer,dpa2}, enhance both accuracy and generalization through message passing\supercite{MPNN} or attention\supercite{shi2022benchmarking}.
However, their increased complexity introduces computational bottlenecks, limiting speed and scalability.

In this work, we build on previous successful models and aim to offer a good trade-off for polymer simulations.
Recent advanced edge-based local GNNs, such as Allegro\supercite{Musaelian2023Learning}, offer a promising approach to addressing the GAS trilemma.
By utilizing edge message passing within localized environments rather than across nodes, Allegro enables efficient parallelization across multiple devices, significantly improving simulation speed while maintaining accuracy and generalization.
However, Allegro's performance is constrained by relatively short cutoffs ($\leq$ \qty{5}{\angstrom}), making it difficult to capture the mid- or long-range atomic interactions prevalent in polymer systems.

\begin{figure}[t]
    \centering
    \includegraphics{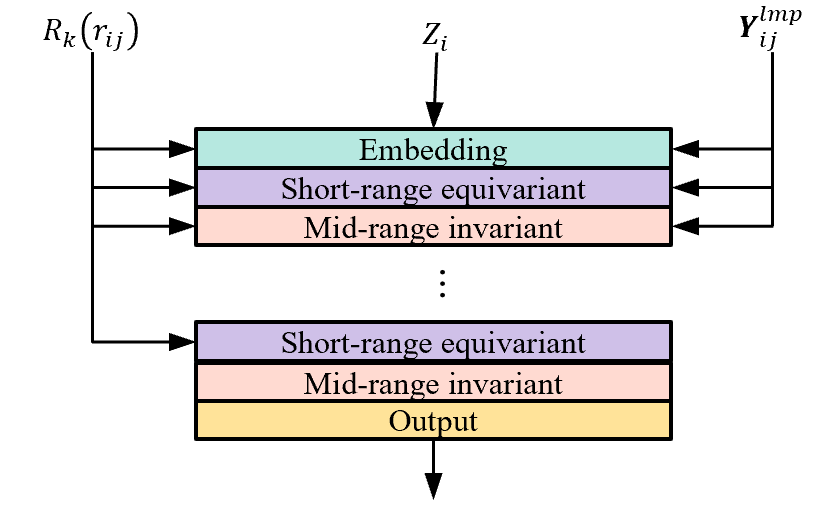}
    \caption{Schematic overview of Vivace's architecture.}
    \label{supfig:model_arch}
\end{figure}

\begin{algorithm}[H]
    \caption{Main forward function}
    \label{supalg:model_arch}
    \begin{algorithmic}[1]
        \Function{MainForward}{$\mZ$, $\mX$, $\rcutmid$, $\rcutshort$, $\lmax$, $\nlayer$, $\nfeat$}
        \State edge\_index = ComputePairEdgeData($\mX$, $\rcutmid$)
        \State $\vh^{(0)}$, $\ve^{(0)}$, $\vt^{(0)}$, $\xi^{(0)}$ = InitialEmbedding($\mX$, $\mZ$, edge\_index, $\nfeat$, $\lmax$)
        \For{$n=0$ to $\nprelayer-1$}
        \State $\vh^{(n+1)}$, $\ve^{(n+1)}$
        = InvariantInteraction(
        $\vh^{(n)}$, $\ve^{(n)}$
        )
        \EndFor
        \For{$n=0$ to $\nlayer-1$}
        \State $\vh^{(n+1)}$, $\ve^{(n+1)}$,$\vxi^{(n+1)}$
        = EquivariantInteraction(
        $\vh^{(n)}$, $\ve^{(n)}$,$\vxi^{(n)}$
        )
        \For{$m=0$ to $\nsub$-1}
        \State $\vh^{(n+1)}$, $\ve^{(n+1)}$
        = InvariantInteraction(
        $\vh^{(n+1)}$, $\ve^{(n+1)}$
        )
        \EndFor
        \EndFor
        \For{$n=0$ to $\npostlayer-1$}
        \State $\vh^{(n+1)}$, $\ve^{(n+1)}$
        = InvariantInteraction(
        $\vh^{(n)}$, $\ve^{(n)}$
        )
        \EndFor
        \State $E$= Output($\vh^\text{final}$, $\ve^\text{final}$)
        \State $q$= \MLP($\vh^\text{final}$)
        \State $E$ += $\sum_{s}\MLP(\ve^\text{final})/r^s$
        \State $\vf = -\frac{\partial E}{\partial \mX}$
        \State $\msigma = \frac{\partial E}{V \partial \mvarepsilon}$
        \State 
        \Return{}  $E$, $\vf$, $\msigma$, $q$
        \EndFunction
    \end{algorithmic}
\end{algorithm}

To overcome these limitations, we introduce Vivace, a new localized model that leverages a mixed-range and mixed invariant-equivariant interaction mechanism to better capture intricate atomic interactions (Alg.~\ref{supalg:model_arch} and Fig.~\ref{supfig:model_arch}).
Our approach builds on the strengths of several successful model architectures, including SchNet\supercite{Schutt2017Quantum}, PaiNN\supercite{schutt2021equivariant}, NequIP\supercite{Batzner2022Equivariant}, Allegro\supercite{Musaelian2023Learning}, TensorNet\supercite{simeon2023tensornet}, and Graphormer\supercite{shi2022benchmarking}.
While maintaining similarity to these prior models, our Vivace architecture introduces key designs for enhanced expressivity and efficiency.
In the following sections, these changes and designs are presented and discussed.
A list of variables can be found in Supplementary Table \ref{suptab:vivace_variable_list}.

\begin{longtable}{@{}p{.15\textwidth}p{.20\textwidth}p{.65\textwidth}@{}}
    \caption{List of Variables used in Vivace}
    \label{suptab:vivace_variable_list}                                                                                                                                                         \\
    \toprule
    \textbf{Variable}      & \textbf{Shape / Type}                   & \textbf{Description}                                                                                                     \\
    \midrule
    \endfirsthead

    \endhead

    \endfoot

    \bottomrule
    \endlastfoot

    $i$                    & $\sZ$                                   & Atom index                                                                                                               \\
    $\natom$               & $\sZ$                                   & Number of atoms in the configurations                                                                                    \\
    $\bx_i$                & $\sR^{3}$                               & Atomic position of atom $i$                                                                                              \\
    $\mX$                  & $\sR^{\natom \times 3}$                 & All atomic positions                                                                                                     \\
    $Z_i$                  & $\sZ$                                   & Atomic species of atom $i$                                                                                               \\
    $\mZ$                  & $\sZ^{\natom}$                          & All atomic species                                                                                                       \\
    \midrule
    $(i, j)$               &                                         & Directed edge with the receiver $i$ and the sender $j$                                                                   \\
    $\localenv(i)$         &                                         & Local environment centered on the atom $i$                                                                               \\
    $\rcutshort$           & $\sR$                                   & Short-range cutoff, \qty{3.8}{\angstrom}                                                                                 \\
    $\rcutmid$             & $\sR$                                   & Mid-range cutoff, \qty{6.5}{\angstrom}                                                                                   \\
    $\rij$                 & $\sR$                                   & Distance between atom $i$ and atom $j$                                                                                   \\
    $\brij$                & $\sR^{3}$                               & Directed edge vector connecting the receiver atom $i$ and the sender atom $j$                                            \\
    $\nedge$               & $\sZ$                                   & Number of edges in the configurations                                                                                    \\
    $\mathrm{edge\_index}$ & $\sZ^{(\nedge,2)}$                      & All directed edges                                                                                                       \\
    \midrule
    $\nfeat$               & $\sZ$                                   & Number of features                                                                                                       \\
    $\nbasis$              & $\sZ$                                   & Number of basis functions                                                                                                \\
    $\nchn$                & $\sZ$                                   & Number of channels in equivariant features                                                                               \\
    $L$                    & $\sZ$                                   & Maximum angular momentum                                                                                                 \\

    $\nprelayer$           & $\sZ$                                   & number of invariant layers before the main chunk                                                                         \\
    $\nsub$                & $\sZ$                                   & number of invariant layers per main layer in the main chunk                                                              \\
    $\npostlayer$          & $\sZ$                                   & number of invariant layers after the main chunk                                                                          \\
    $\nlayer$              & $\sZ$                                   & number of layers in the main chunk
    \\
    \midrule
    $n$                    & $\sZ$                                   & Layer index                                                                                                              \\
    $l$                    & $\sZ$                                   & Angular momentum quantum number                                                                                          \\
    $m$                    & $\sZ$                                   & Magnetic quantum number                                                                                                  \\
    $p$                    & $\sZ$                                   & Parity                                                                                                                   \\
    $k$                    & $\sZ$                                   & Component index                                                                                                          \\
    $s$                    & $\sZ$                                   & Power index for energy modification term                                                                                 \\
    \midrule
    $\rho_{ij}$            & $\sR^{\nchn}$                           & Edge length embedding                                                                                                    \\
    $Y^{lmp}_{ij}$         & $\sR$                                   & Real spherical harmonics                                                                                                 \\
    $\vt_{ij}$             & $\sR^{\nfeat}$                          & Edge type encoding                                                                                                       \\
    $\tilde{\mY}_{ij}$     & $\sR^{(L+1)^2}$                         & Real spherical harmonics (alternative notation)                                                                          \\
    $\ve_{ij}^{(n)}$       & $\sR^{\nfeat}$                          & Edge invariant feature at layer $n$                                                                                      \\
    $\vh^{(n)}_i$          & $\sR^{\nfeat}$                          & Invariant node feature of atom $i$ at layer $n$                                                                          \\
    $\vxi^{lmp,(n)}_i$     & $\sR^{\nchn}$                           & Equivariant node feature of atom $i$ with angular momentum $l$, magnetic quantum number $m$ and parity $p$, at layer $n$ \\
    $\txi_i^{(n)}$         & $\sR^{\nchn \times (L+1)^2}$            & Equivariant node features (alternative notation) at layer $n$                                                            \\
    \midrule
    $\rva_i$               & $\sR^{\nfeat}$                          & Initial atomic encoding of atom $i$                                                                                      \\
    $\rvb_{ij}$            & $\sR^{\nbasis}$                         & Radial basis function encoding of edge $(i,j)$                                                                           \\
    $\mu_k$                & $\sR$                                   & Mean value for radial basis function $k$                                                                                 \\
    $\mW_k$                & $\sR^{\nfeat \times \text{n\_species}}$ & Weight matrix for atomic encoding                                                                                        \\
    $\delta$               & $\sR$                                   & Smoothing parameter for cutoff function                                                                                  \\
    $\fcut(x)$             & $\sR$                                   & Smooth cutoff/envelope function                                                                                          \\
    $\rmod$                & $\sR$                                   & Modification cutoff radius (\qty{1.5}{\angstrom})                                                                        \\
    \midrule
    $\vq_i$                & $\sR^{\nfeat}$                          & Query vector for attention mechanism                                                                                     \\
    $\vk_{ij}$             & $\sR^{\nfeat}$                          & Key vector for attention mechanism                                                                                       \\
    $\vv_{ij}$             & $\sR^{\nfeat}$                          & Value vector for attention mechanism                                                                                     \\
    $\valpha_{ij}$         & $\sR$                                   & Attention weights                                                                                                        \\
    $\vw_{ij}^{lp}$        & $\sR^{\nchn}$                           & Learnable weights for equivariant aggregation for angular momentum $l$ and parity $p$                                    \\
    $d$                    & $\sZ$                                   & Feature dimension for attention scaling                                                                                  \\
    $\text{Norm}$          & $\sR$                                   & Normalization factor for tensor product                                                                                  \\
    \midrule
    $E$                    & $\sR$                                   & Total energy                                                                                                             \\
    $E_i^{\text{node}}$    & $\sR$                                   & Node energy contribution                                                                                                 \\
    $E_{ij}^{\text{edge}}$ & $\sR$                                   & Edge energy contribution                                                                                                 \\
    $E_{ij}^{\text{mod}}$  & $\sR$                                   & Short-range energy modification                                                                                          \\
    $E_{Z_i}$              & $\sR$                                   & Reference energy for atomic species $Z_i$                                                                                \\
    $\vf_i$                & $\sR^3$                                 & Force on atom $i$                                                                                                        \\
    $\msigma$              & $\sR^{3 \times 3}$                      & Stress tensor                                                                                                            \\
    $\mSigma$              & $\sR^{3 \times 3}$                      & Stress tensor (alternative notation)                                                                                     \\
    $\mvarepsilon$         & $\sR^{3 \times 3}$                      & Lattice strain tensor                                                                                                    \\
    $V$                    & $\sR$                                   & Lattice volume                                                                                                           \\
\end{longtable}

\subsection{Local environment construction\label{sec:localenv}}

For each atom, its local environment $\localenv(i)$ is defined as all directed edges $(i\leftarrow j)$ between itself and its neighboring atoms within a cutoff radius $\rcut$.

\begin{equation}
    \label{eq:local_env}
    \localenv_{\rcut}(i) = \{i \leftarrow j \mid r_{ij} \leq \rcut\}.
\end{equation}
The interactions between the central atom and its neighbors are restricted to these local environments.
This means the local environments do not overlap, i.e., $\localenv(i) \cap \localenv(j) = \emptyset$ for $i \neq j$.
For simplicity, we write $\sum_{j \in \localenv (i)}$ to mean $\sum_{i \leftarrow j \in \localenv(i)}$.

Within our model, two local environments, $\localshort$ and $\localmid$, are used for each atom, defined by a short-range cutoff $\rcutshort$ and a mid-range cutoff $\rcutmid$, respectively.
Equivariant features and tensor products are limited to the short-range local environment, which contains fewer edges due to the shorter cutoff, making the model less computationally intensive.
We set $\rcutshort=\qty{3.8}{\angstrom}$ and $\rcutmid=\qty{6.5}{\angstrom}$.
In addition, we apply a polynomial energy modification to edges shorter than $\rmod=$ \qty{1.5}{\angstrom}.

\subsection{The initial embedding layer}

\subsubsection{Encoding and envelope function}
\begin{figure}
    \centering
    \includegraphics{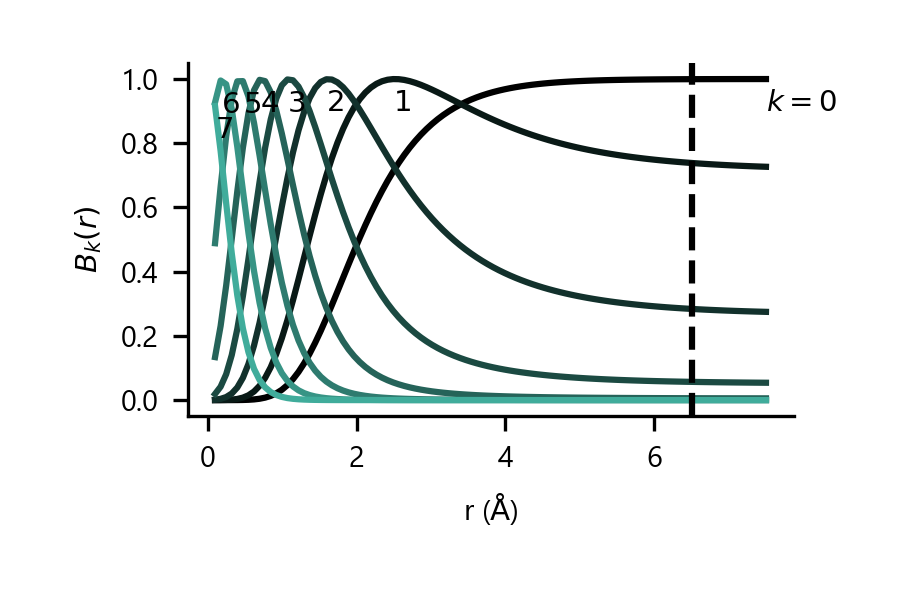}
    \caption{Examples of exponential basic function with $\nbasis=8$, $\rmax=\qty{6.5}{\angstrom}$. The value of different $k$ is labeled by text next to the maximum value of each line.}
    \label{supfig:expnorm}
\end{figure}

We first encode each atom node $i$ with atomic species $Z_i$ to $\rva_i \in \sR^{\nfeat}$ using one-hot encoding and a linear transformation,
\begin{equation}
    \rva_i^k = \mW_k \cdot \left( \text{1Hot}\left( Z_i \right) \right),
\end{equation}
and each edge $(i\leftarrow j)$ with length $\rij$ to $\rvb_{ij} \in \sR^{\nbasis}$  with exponential radial basis functions \supercite{Unke2019PhysNet,simeon2023tensornet} (Fig.\ref{supfig:expnorm})
\begin{equation}
    \rvb_{ij}^k = B_k(\rij) = \exp \left [ -\frac{\nbasis^2}{4(1-e^{-\rcut})^2} \left( e^{ -\rij}  - \mu_k \right)^2 \right ],
\end{equation}
where $k$ is the $k$-th component of these encoding vectors and the mean value is
\begin{equation}
    \mu_k = \frac{k}{\nbasis} + \left(1-\frac{k}{\nbasis}\right)e^{-\rcut}.
\end{equation}

For each directed edge $(i\leftarrow j)$, the edge type can also be encoded with an MLP,
\begin{equation}
    \vt_{ij} = \text{MLP}\left( \va_i \concat \va_j \right),
\end{equation}
where $\concat$ denotes concatenation.
This quantity is later used in mid-range interactions (Eq. \eqref{eq:use_edge_type})

\begin{figure}
    \centering
    \includegraphics{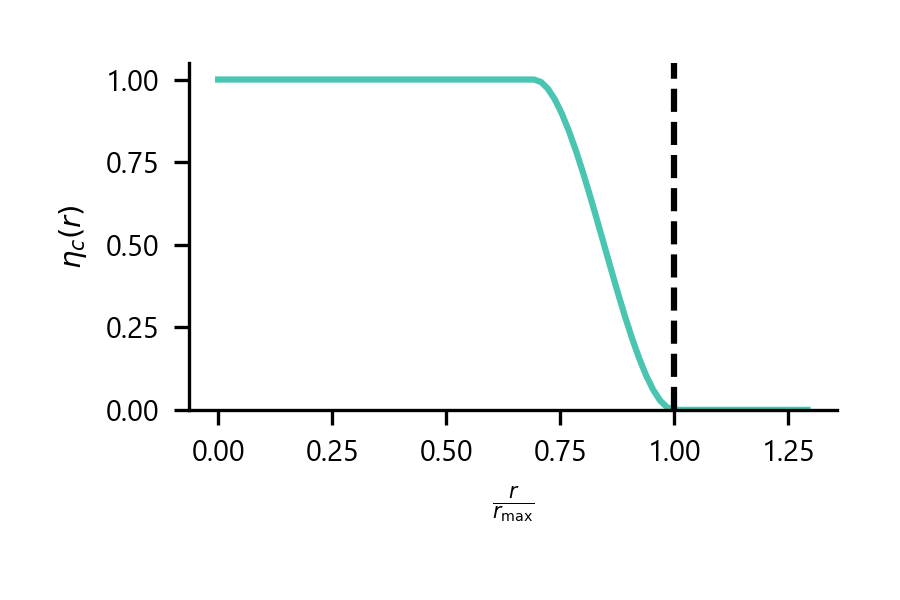}
    \caption{Example function value of smooth step function $\fcut$. $\delta=\frac{2}{6.5}$}
    \label{supfig:cutoff}
\end{figure}

Because the contribution of each atom pair to the total energy should smoothly go to zero as their distance approaches the cutoff radius, a smooth cutoff function is often used.
In this work, we use a smooth step function $\fcut(x)$ where $x = r_{ij}/r_\text{cut}$ (Fig. \ref{supfig:cutoff}),
\begin{equation}
    \fcut(x) =
    \begin{cases}
        1                                         & \text{if } 0 \le x \le 1 - \delta \\
        3\left(1-x\right)^2 - 2\left(1-x\right)^3 & \text{if } 1 - \delta < x < 1     \\
        0                                         & \text{if } x \ge 1
    \end{cases}\, ,
\end{equation}
where $\delta$ is a small parameter that controls the smoothing region.
We note that many previous models\supercite{Pelaez2024,simeon2023tensornet} use a cosine cutoff function, which usually works well with relatively small cutoffs, such as \qtyrange{4.5}{5}{\angstrom}.
However, the cosine function can introduce a strong inductive bias and prevent the model from learning interactions at larger distances.

\subsubsection{Initialization of edge-related features}

For each directed edge $(i,j)$, there are three associated quantities: the edge type $\rvt_{ij}$, the edge length embedding $\rho(\rij)$, and the edge invariant feature $\ve_{ij}$.
The edge length embedding $\rho(\rij)$ depends only on the length of this edge, while $\ve_{ij}$ is updated in each interaction block with information about atom species and other edges in the same local environment.

The edge length feature is a simple transform from $\rvb_{ij}$,
\begin{equation}
    \rho_{ij}^{\short, \medium} = \MLP \left ( \rvb^{\short, \medium}_{ij} \right )\fcut\left (\frac{\rij}{\rcut^{\short, \medium}} \right ).
\end{equation}
Depending on whether this embedding is used in short- or mid-range interactions, the envelope and the basis set function will use the corresponding cutoff radius.

Initial edge invariant features $\ve^{(0)}_{ij} \in \sR^{ \nfeat}$ are a function of both distance and species information,
\begin{equation}
    \ve^{(0)}_{ij} = \text{MLP}\left( \rvb_{ij}^{\medium} \concat \va_i \concat \va_j \right) \fcut(r_{ij}/\rcutmid).
\end{equation}

\subsubsection{Initialization of node-related features}

For each atom $i$, there are two quantities: invariant node features $\vh_i$ and equivariant node features $\tilde{\vxi}_i$.
Both are updated in later parts of the model.

The initial invariant node features $\vh^{(0)}_i$ are simply the atomic specie encoding,
\begin{equation}
    \vh^{(0)}_i = \rva_i.
\end{equation}

Initial equivariant node features $\tilde{\vxi}^{(0)}_{i} \in \sR^{ \nchn \times (L+1)^2}$ are computed through a learnable aggregated embedding of edge real spherical harmonics $Y^{lm}_{ij}$,
\begin{align}
    \vxi^{(0),lm}_{i} & = \sum_{j \in \localshort(i)}\vw^{(0),l}_{ij} Y^{lm}_{ij},
    \\
    \vw^{(0),l}_{ij}  & = \text{MLP}\left( \rvb_{ij}^\short \concat \vh^{(0)}_i \concat \vh^{(0)}_j \right) \fcut(r_{ij}/\rcutshort).
\end{align}
Here $n$ denotes the channel index, $l$ represents the angular momentum order from 0 to $L$, and $m$ labels the magnetic quantum number within a specific $l$, ranging from $-l$ to $l$.
$Y^{lm}_{ij} \in \sR$ is the spherical harmonic component of the normalized relative atomic position $\hat{\vr}_{ij} = \vr_{ij}/|\vr_{ij}|$.

\subsection{Mid-range invariant interaction \label{section:midrange}}
In this interaction block, only node and edge invariant features are used as input and are updated.
\begin{align}
    \ve^{(n+1)}_{ij} & = \ve_{ij}^{(n)} + \MLP(\ve_{ij}^{(n)})\MLP(\rvt_{ij}),
    \label{eq:use_edge_type}
    \\
    \vh^{(n+1)}_i    & = \vh_i^{(n)}
    + \MLP \left [
        \text{LayerNorm}
        \left (
        \vh_i^{(n)}+\Delta \vh_i
        \right )
        \right],
    \\
    \Delta \vh_{i}   & = \text{Linear} \left ( \sum_{j \in \localmid(i) } \valpha_{ij} \vv_{ij} \fcut \left( r_{ij}/\rcutmid \right) \right ),
\end{align}
where $n$ denotes the $n$-th layer, and intermediate quantities are computed as
\begin{align}
    \brho_{ij}   & = \text{MLP}(\brho^\medium_{ij}),
    \\
    \vq_i        & = \text{Linear}\left( \vh_i \right),
    \\
    \vk_{ij}     & = \text{Linear}\left( \ve^{(n+1)}_{ij} \right) + \brho_{ij},
    \\
    \vv_{ij}     & = \text{Linear}\left( \ve^{(n+1)}_{ij} \right) + \brho_{ij},
    \\
    \valpha_{ij} & = \exp \left(
    - \frac{\left\| \vq_i - \vk_{ij} \right\|^2_2}{2 \sqrt{d}}
    \right) \fcut \left( r_{ij}/\rcutmid \right).
    \label{eq:our-attention}
\end{align}

\begin{figure}[H]
    \centering
    \includegraphics[width=0.8\textwidth]{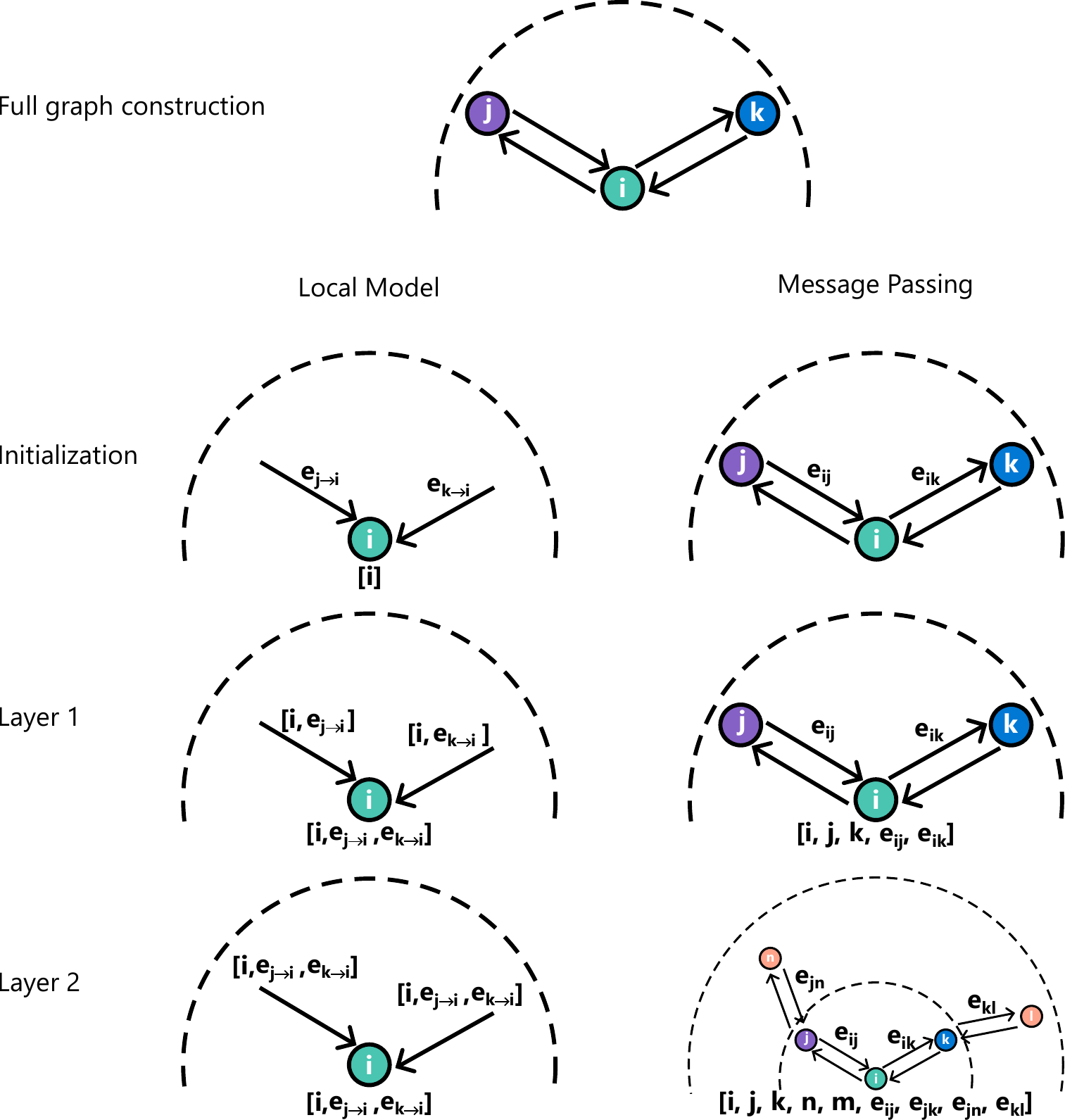}
    \caption{A comparison of information flow in local (left) and message passing(right) model.
        The diagram illustrates how the information available to a central node $i$ evolves through successive layers. (Left) In a local model, like Vivace, the update of node $i$ only ever uses information from its initial 1-hop neighborhood. The receptive field remains fixed, though the features become more complex. (Right) In a conventional message passing model, each layer aggregates updated information from its neighbors. This allows the receptive field of node $i$ to expand with each layer, incorporating information from nodes two hops away ($m$, $n$) after the second layer.
        \label{fig:message_passing}
    }
\end{figure}

At first glance, our design appears similar to a typical self-attention block in a graph transformer \supercite{GeoMFormer},
which uses features of the center node ($\vh_i$), neighbor node ($\vh_j$), and their connecting edge ($\vr_{ij}$) to derive a query $\vq_i$, key $\vk_j$, attention bias $\vb_{ij}$ and value $\vv_j$:
\begin{align}
    \vh_i^{(n+1)} & = \vh_i^{(n)} + \text{softmax}_{j \in \fullenv(i)}(\vq_i \cdot \vk_j / \sqrt{d}+\vb_{ij}) \vv_j.
    \label{eq:graph-attention}
\end{align}
This mechanism (Eq. \eqref{eq:graph-attention}) is a form of message passing, which can be interpreted as
\begin{equation}
    \vh_i^{(n+1)}  = \vh_i^{(n)} + \sum_{j} f \left (
    \vh_i^{(n)}, \vh_j^{(n)}, \vr_{ij}
    \right ).
    \label{eq:message-passing-sup}
\end{equation}
In this scheme, each update layer incorporates the updated states of neighboring atoms to the central atom.
This causes the receptive field to expand with each update layer, allowing information to propagate from atoms farther away:
\begin{align}
    \begin{split}
        \vh_i^{(0)} & \leftarrow \fullenv(i),
        \\
        \vh_i^{(1)} & \leftarrow \fullenv(i),~\fullenv(j) \mid j \in
        \fullenv (i),
        \\
        \vh_i^{(2)} & \leftarrow \fullenv(i),~\fullenv(j) \mid j \in
        \fullenv (i),~\fullenv (k) \mid k \in \fullenv \left ( \fullenv (i) \right ).
    \end{split}
\end{align}
This propagation is visualized on the right in Fig. \ref{fig:message_passing}.
When using the model over multiple GPUs, message passing can occur across GPUs, which can cause synchronization blocking (happening $2\nlayer$ times for a $\nlayer$-layer model) and is non-trivial to implement.

However, a subtle difference in how information is sourced leads to a substantial difference in the model's receptive field.
Our update mechanism (Eq. \eqref{eq:our-attention}) is a local interaction, which can be summarized as
\begin{equation}
    \vh_i^{(n+1)}  = \vh_i^{(n)} + \sum_{j} f \left (
    \vh_i^{(n)}, \ve_{ij}^{(n)}
    \right )
    \label{eq:local-interaction-sup}.
\end{equation}
As shown on the left in Fig. \ref{fig:message_passing}, no matter how many update layers are applied, the features for node $i$ ($\vh_i$) are always derived exclusively from its initial 1-hop local environment ($\localenv(i)$).
Ultimately, this distinction highlights a fundamental trade-off: local models offer higher computational efficiency by restricting information flow, whereas message-passing models achieve potentially greater expressive power by systematically expanding their receptive field.

In addition to this local constraint, our choice of the kernel function in Eq. \eqref{eq:our-attention} also diverges from typical options.
Conventional attention mechanisms use a scaled dot product and a softmax function to compute attention scores.
In the MLFF landscape, other variants exist.
For example, TorchMD-Net \supercite{tholke2022torchmd, Pelaez2024} applies the SiLU activation function to the dot product:
\begin{equation}
    \valpha_{ij} = \text{SiLu}(\vq_i \cdot \vk_j) \fcut \left( r_{ij}/\rcutmid \right).
\end{equation}
Another common strategy, often employed in message-passing models, replaces the dot-product interaction with a more general function.
This function typically uses an MLP that operates on the concatenated features of the interacting pair:
\begin{equation}
    \valpha_{ij} = \MLP(\vq_i \concat \vk_j) \fcut \left( r_{ij}/\rcutmid \right).
\end{equation}
We considered and tested all these options during our hyper-parameter search.
Ultimately, the Gaussian distance kernel\supercite{SOFT} used in Eq.~\eqref{eq:our-attention} and softmax were found to yield the best performance.

\subsection{Short-range SE(3)-equivariant interaction\label{sec:short-equiv}}

This block updates equivariant node features and invariant edge features, using only invariant features as input.

The core interactions in this block are defined as
\begin{align}
    \txi^{(n+1)}_{i} & = \txi_{i}^{(n)} +
    \Delta \txi_{i}
    \\
    \ve_{ij}^{(n+1)} & = \text{MLP} \left ( \Delta \ve_{ij}  + \ve_{ij}^{(n)} \right ) + \ve_{ij}^{(n)},
    \\
    \Delta \txi_{i}  & =
    \frac{1}{\text{Norm}}  \tilde{\vxi}_{i}^{(n)}
    \bigotimes
    \sum_{j\in\localshort(i)}\vw_{ij} \tilde{Y}_{ij},
    \label{eq:tensorproduct-sup}
    \\
    \Delta \ve_{ij}  & = \text{Linear} \left ( \Delta \tilde{\vxi}_{i}\cdot \tilde{Y}_{ij} \right )  \MLP(\vrho_{ij}),
    \label{eq:dotproduct-sup}
\end{align}
where $\bigotimes$ is the tensor product operation, which will be detailed in the next section.
Intermediate quantities are computed as follows:
\begin{align}
    \brho_{ij}    & = \text{MLP}(\brho^\short_{ij}),
    \\
    \vq_i         & = \text{Linear}\left( \vh_i \right),
    \\
    \vk_{ij}      & = \text{Linear}\left( \ve_{ij} \right),
    \\
    \vv_{ij}      & = \text{Linear}\left( \ve_{ij} \right),
    \\
    \valpha_{ij}  & = \exp \left(
    - \frac{\left\| \vq_i - \vk_{ij} \right\|^2_2}{2 \sqrt{d}}
    \right) \fcut \left( r_{ij}/\rcutshort \right),
    \\
    \vw^{lp}_{ij} & = \text{Linear}_{lp}\left( \valpha_{ij} \vv_{ij} \right).
\end{align}
The computation of the normalization factor $\text{Norm}$ in Eq. \eqref{eq:tensorproduct-sup} is detailed in Alg. \ref{supalg:equivariant_update}.

This interaction can be interpreted as a form of \emph{cross-attention} in the context of transformer models like DPA-2 \supercite{dpa2} and GeoMFormer \supercite{GeoMFormer}.
Here, the attention weights $\valpha_{ij}$ are computed from invariant quantities but are applied to an equivariant quantity.
While DPA-2 and GeoMFormer implement four types of attention mechanisms—(1) self-attention on invariant features, (2) cross-attention from invariant to equivariant features, (3) cross-attention from equivariant to invariant features, and (4) self-attention on equivariant features—Vivace includes only types (1) and (2).

This interaction block is also similar to Allegro \supercite{Musaelian2023Learning}, which uses a summation over the local environment, $\sum_{j\in\localshort(i)}\vw^{l}_{ij} Y^{lm}_{ij}$, as the right-hand side of its tensor product.
A key difference is that Allegro updates equivariant edge features, leading to significantly higher memory consumption:
\begin{equation}
    \Delta \tilde{\vxi}_{ij} = \tilde{\vxi}_{ij} \bigotimes
    \sum_{j\in\localshort(i)}\vw_{ij} \tilde{Y}_{ij}.
    \label{eq:allegro}
\end{equation}
The complexity of this tensor product is $O(\nedge)$, whereas in Vivace, it is only $O(\natom)$ because it updates node features.

The most distinctive feature of Vivace is its inner product in Eq.~\eqref{eq:dotproduct-sup}.
By substituting Eq.~\eqref{eq:tensorproduct-sup} into Eq.~\eqref{eq:dotproduct-sup}, we get:
\begin{align}
    \begin{split}
        \Delta \ve_{ij}
         & =  \left [\txi_{i}
            \bigotimes
            (\sum_{k\in\localshort(i)}\vw_{ik} \tilde{Y}_{ik})
            \cdot \tilde{Y}_{ij} \right ]
        \MLP(\rho_{ij}).
    \end{split}
\end{align}
We can reorder the equation as follows:
\begin{align}
    \begin{split}
        \Delta \ve_{ij}
         & \propto
        \sum_{k\in\localshort(i)}
        \left [
            \left (
            \txi_{i}
            \bigotimes
            (\vw_{ik} \tilde{Y}_{ik})
            \cdot \tilde{Y}_{ij}
            \right )
            \MLP(\rho_{ij})
            \right ]
        \\
         & =
        \sum_{k\in\localshort(i)}
        \tilde{\mA}_{ik} \cdot \tilde{\mB}_{ij},
    \end{split}
\end{align}
where
\begin{align}
    \begin{split}
        \tilde{\mA}_{ik} & = \tilde{\vxi}_{i} \bigotimes \left ( \vw_{ik}\tilde{Y}_{ik} \right ),
        \\
        \tilde{\mB}_{ij} & = \tilde{Y}_{ij} \MLP(\rho_{ij}).
    \end{split}
\end{align}
This invariant edge feature update is equivalent to a three-body interaction term \supercite{M3GNet} or a triangular attention mechanism\supercite{abramson2024accurate}, which typically takes the form:
\begin{equation}
    \Delta \ve_{ij} = \sum_{k} \valpha_{ijk} \vv_{ik}.
\end{equation}
For a central atom with $M$ neighbors, a standard three-body interaction scales as $O(M^2)$ due to the double summation over indices $j$ and $k$.
In contrast, our update method maintains a computational complexity of only $O(M)$ because there is only a single summation in Eq.~\eqref{eq:tensorproduct-sup}.

\subsubsection{Tensor Product\label{sec:tp}}

The tensor product operation allows different angular momentum channels to interact while preserving equivariance.
Let us consider the full tensor product of two SE(3)-equivariant tensors in an irreducible representation:
\begin{align}
     & \tilde{\rva}^{l_1}  \bigotimes \tilde{\rvb}^{l_2}  = \bigoplus_{|l_1-l_2| \leq l_3 \leq l_1+l_2} \tilde{\rvc}^{l_3}.
\end{align}
We can explicitly write out how each component interacts with each other:
\begin{align}
    c^{n_3l_3m_3} & =
    w^{l_3 \leftarrow l_1 l_2}_{n_1 n_2 n_3}
    \sum_{m_1, m_2}
    C_{(l_1 m_1),(l_2 m_2)}^{(l_3 m_3)}
    \left( a^{n_1l_1m_2}
    b^{n_2l_2m_2}
    \right),
    \label{eq:path_tp}
\end{align}
where $n_{1, 2, 3}$ are the channel-wise dimensions.
From this equation, we can see that the complexity of the tensor product is dictated by the angular momenta $l_{1,2}$ and the size of the weight matrix.
If we allow all possible combinations of $n_{1,2,3}$, the complexity scales as $O(\nchn^3)$.
If $\tilde{\rva}$ and $\tilde{\rvb}$ both have components from $l=0$ to $\lmax$, the weight matrix has an additional factor of $O(\lmax^3)$.

In Vivace, we only use the simplest tensor product.
We assume that the weight matrix in Eq.~\eqref{eq:path_tp} can be decomposed as:
\begin{align}
    w^{l_3 \leftarrow l_1 l_2}_{n_1n_2n_3} = U^{l_1}_{n_1} V^{l_2}_{n_2} Q^{l_3}_{n_3}.
\end{align}
In this way, the total number of learnable weights in the system depends only linearly on $\lmax$.
The matrices $U$, $V$, and $Q$ can then be absorbed into the values of $\rva$, $\rvb$, and $\rvc$:

\begin{align}
    \tilde{\vc}^{l_3} =
    \mQ^{l_3}
    \left(
    \left (\mU^{l_1} \tilde{\va}^{l_1} \right )
    \otimes
    \left ( \mV^{l_2} \tilde{\vb}^{l_2} \right )
    \right)^{l_3}.
\end{align}
For the case of a "uuu" operation (assuming $n_1 = n_2 = n_3$), where tensor products are only applied to the same channel, the tensor product can be further simplified to a tri-matrix product:
\begin{align}
    c^{nl_3m_3} =
    \sum_{l_3 \leftarrow l_1 l_2}
    \sum_{m_1} \sum_{m_2}
    C_{(l_1 m_1),(l_2 m_2)}^{(l_3 m_3)}
    a^{nl_1m_1}
    b^{nl_2m_2}.
    \label{eq:ltp}
\end{align}

Compared to the original form of the tensor product in Eq.~\eqref{eq:path_tp}, this simplified tensor product may be less expressive\supercite{xie2025price}, but its complexity is reduced to $O(\nchn)$, and the associated weight matrix scales as $O(\lmax)$.
This simplest form of tensor product reduces the memory footprint and is relatively easy to implement.

\begin{algorithm}
    \caption{Equivariant Interaction}
    \label{supalg:equivariant_update}
    \begin{algorithmic}[1]
        \Function{EquivariantInteraction}{$\vh$, $\ve$, $\vxi$, $\fcut^{\short}$, $\tilde{Y}$}
        \State $Q=$ Linear($\vh$)[Receiver]
        \State $K=$ Linear($\ve$)
        \State $V=$ Linear($\ve$)
        \State $A=$ AttentionType($Q$, $K$, $V$, $\fcut^{\short}$, $\vrho$)
        \State $A=$ Linear($A$)
        \State $\vxi_\text{temp}=$ Reshape(A) $\tilde{Y}$
        \State local\_env = scatter\_sum($\vxi_\text{temp}$, index=receiver)
        \State $\vxi_\text{new}$ = TensorProduct($\vxi$, local\_env)
        \State $\text{Norm}$ = Sum(($\vxi_\text{new}$)$^2$, dim=-2)
        \Comment{sum over all $l, m$}
        \State $\text{Norm}$ = Mean($N$, dim=-1)
        \Comment{take mean over channels}
        \State $\vxi_\text{new} = \vxi_\text{new} /( \text{Norm}+10^{-5}) $
        \If{update\_edge}
        \State tmp\_edge = $\vxi_\text{new}$[receiver]
        \State $\ve_\text{new}$ = InnerProduct(tmp\_edge, $\tilde{Y}$)
        \State edge\_filter = MLP($\vrho$)
        \State $\ve_\text{new}$ =  MLP($\ve_\text{new}\times$ edge\_filter + $\ve$)
        \State $\ve_\text{new}$ =  $\ve_\text{new}\times \fcut^\short$  + $\ve$
        \Else
        \State $\vh_\text{new}$ = $\vxi_\text{new}$[l=0]
        \State $\vh_\text{new}$ = MLP(LayerNorm(MLP($\vh_\text{new}$))) + $\vh$
        \EndIf
        \State 
        \Return{} $\vh_\text{new}$, $\ve_\text{new}$, $\vxi_\text{new}$
        \EndFunction
    \end{algorithmic}
\end{algorithm}

\subsection{Output layer}
In Vivace, the total energy is a sum of contributions from nodes, edges, and atomic reference energies.
The energy of each node, $E_i^{\text{node}}$, and edge, $E_{ij}^{\text{edge}}$, is computed by passing their final features through an MLP without biases in all its linear layers.
An additional energy modification term, $E_{ij}^{\text{mod}}$, is used to address the strong repulsion and attraction when two atoms are too close.
Instead of using a ZBL potential, we simply add a polynomial correction in Eq.~\eqref{eq:zbl-mock}.
\begin{align}
    E_i^{\text{node}}    & = \text{MLP}(\vh_i^{\text{final}}),
    \\
    E_{ij}^{\text{edge}} & = \text{MLP}(\ve_{ij}^{\text{final}}),
    \\
    E_{ij}^{\text{mod}}  & =
    \sum_s
    \frac{1}{\rij^s}\text{MLP}_s(\ve_{ij}^{\text{final}})\fcut(\rij/\rcutmod),
    ~\text{if}~ r_{ij} \leq \qty{1.5}{\angstrom},
    \label{eq:zbl-mock}
\end{align}
The total energy can be written as:
\begin{align}
    E & = \sum_i E_i^{\text{node}} + \sum_{i,j} (E_{ij}^{\text{edge}}+E_{ij}^{\text{mod}}) + \sum_i E_{Z_i},
\end{align}
where $E_{Z_i}$ is the reference energy of atom type $Z_i$.
Forces $\vf_i$ and stresses $\msigma$ are then derived from the total energy with respect to nuclear positions $\vr_i$ and lattice strain $\mvarepsilon$, respectively, where $V$ is the lattice volume.
\begin{align}
    \vf_i & = -\frac{\partial E}{\partial \vr_i},  \quad
    \msigma = \frac{\partial E}{V \partial \mvarepsilon}.
\end{align}

\begin{table}[H]
    \caption{Hyperparameter Search Space for Vivace Model Optimization}
    \label{tab:optuna-params}
    \setlength{\tabcolsep}{0pt}
    \begin{tabular*}{\linewidth}{@{}l@{\hspace{5mm}}l@{\extracolsep{\fill}}p{74mm}@{}}
        \toprule
        Parameter & Distribution  & Description \\
        \midrule
        $\lmax$ & $ \Unif\{0,\dots,6\}$ & Maximum $L$ on spherical harmonics\\
        $\nlayer$ & $ \Unif\{0, \dots, 6\}$ & \# of layers in main chunk\\
        $\nprelayer $ & $ \Unif\{0, 1, 2\}$ & \# of invariant layers before main chunk\\
        $\npostlayer $ & $ \Unif\{0, 1, 2\}$ & \# of invariant layers after main chunk\\
        $\nsub $ & $ \Unif\{0, 1\}$ & \# of invariant interaction block per main chunk layer\\
        $\mlpd$ & $ \Unif\{1,2\}$ & \# of hidden layers in MLP in the interaction block\\
        $\nbasis $ & $ \Unif\{8, 12, 16, \dots, 32\}$ & \# of radial basis sets\\
        $\nfeat $ & $ \Unif\{8, 16, 32, \dots, 1024\}$ & \# of feature for invariant components\\
        $n_{\mathrm{attn-heads}} $ & $ \Unif\{4, 8, 16, 32, 64\}$ & \# of attention heads \\
        $\nchn$ & $ \Unif\{8, 16, 32, \dots, 1024\}$ & \# of features for equivariant components \\
        \cmidrule(r){1-1}
        attn type & Choice\{softmax, silu, concat, exp\}  & the kernel function to compute attention $\valpha_{ij}$
        \\
        \cmidrule(r){1-1}
        lr & $\LogUnif(10^{-4}, 0.005)$ & Learning rate\\
        Epochs & 5 & Number of training epochs \\
        Batch size & 2400 & Batch size during training \\
        \bottomrule
    \end{tabular*}
\end{table}

\section{Model Training}
\label{sec:hyperopt}

As shown in Section~\ref{sec:model_details}, Vivace has tons of hyperparameters and choices of functionals for the same architecture.
Thus, we first focused on comprehensive hyperparameter optimization.
We employed three distinct datasets for this selection process, which served the dual purpose of finding optimal settings and partially mitigating initial training bias.
We used multivariate TPE Sampling\supercite{NIPS2011_86e8f7ab,optuna_2019} to iteratively sample and evaluate model configurations from the hyperparameter space listed in Table~\ref{tab:optuna-params} (upper section).
To evaluate a configuration, we first employ an A100 GPU to measure the MD simulation speed of the untrained model on a \num{2000}-atom system and prune configurations that fail to exceed \qty{0.5}{\nspday}.
We then sample a training configuration (Table \ref{tab:optuna-params}, lower section) and train the model.
During training, we prune unpromising runs using Hyperband\supercite{hyperband2018}.

Over one week of continuous execution, this phase accumulated \num{1000} to \num{2000} distinct model performance results.
The final optimal hyperparameters for the subsequent stages were determined by selecting the configuration that demonstrated the best accuracy on PolyPack and PolyCrop; specifically, we chose the best parameter sets for different simulation speeds.

Table \ref{tab:winning_model} shows a list of winning models found by this hyperparameter optimization process.
These winning model setups are then trained with different training data and training protocols.
Around five different protocols were tested.
The model-protocol combination that yields the best PolyDiss validation set performance is then chosen for the final testing on density and glass transition temperature.

The model presented in the main text is model~4 listed in Table~\ref{tab:winning_model}.
It utilizes single precision (float32) and consists of three equivariant interaction layers with $\lmax=2$, 8 radial basis sets, invariant features of length 64, and equivariant features of length 32.
In total, the model contains \num{389005} parameters.
This model presents the best speed and accuracy trade-off among the six listed in the table.
However, it has $\nsub=0$, indicating purely two-body mid-range interactions.
Further constraint optimization with $\nsub=1$ should significantly improve the model's expressivity, but this is beyond the scope of this work.

This model is trained in two stages.
In the first stage, Omol, PolyCrop-ORCA, and PolyCrop-XTB are used.
In this stage, the model features two heads for energy/force prediction.
One head predicts energy on the $\omega$B97M-V functional and def2-QZVP basis set, while the other head predicts DFT energy with the $\omega$b97x-v functional and def2-TZVPP basis set.
The sampling ratios on these three datasets are 0.05:1:1, respectively.
In the second stage, PolyPack and PolyDiss are included in the training set.
The sampling ratio is 10 (PolyDiss) : 0.2 (Omol) : 1 (PolyPack) : 0.5 (PolyCrop-ORCA) : 0.05 (PolyCrop-XTB).
This protocol results in a final model with three energy prediction heads, predicting two ORCA-calculated energies and one CP2K-calculated energy.

In both stages, the model is trained with the AdamW optimizer\supercite{adamw}.
The learning rate is scheduled to be reduced by a factor of \num{0.9} if the validation loss stays on a plateau for more than \num{200000} batches.
The starting learning rates are \num{0.0001} and \num{0.0005} for the two stages, respectively.
The batch on each GPU is sampled dynamically such that each batch has around \num{4800} atoms from all the graphs in the batch combined.
This model was trained with four A100 GPUs for four days in the first stage and two days in the second stage.

\begin{table}[H]
    \caption{Example models found by Hyper-parameter optimization. All models have $\nprelayer=0$. The speed is measured at A100 with \num{1944}-atom polystyrene configurations with \qty{0.5}{\fs} time step.
    }
    \label{tab:winning_model}
    \centering
    \setlength{\tabcolsep}{0pt}
    \begin{tabular*}{\linewidth}{@{}@{\extracolsep{\fill}}lccccccccccc@{}}
        \toprule
        id & Speed     & $\lmax$ & $\nlayer$ & $\nbasis$ & $n_\text{heads}$ & $\text{attn.}$ & $\nfeat$ & $\mlpd$ & $\nchn$ & $\npostlayer$ & $\nsub$ \\
        & (\unit{\nspday})    &  &&&& type \\
        \midrule
        0   & 0.45     & 4       & 3         & 20        & 8                & exp        & 64       & 1       & 64       & 2             & 1
        \\
        1   & 0.90     & 3       & 3         & 16        & 64               & exp        & 64       & 2       & 64       & 0             & 0
        \\
        2   & 0.85     & 3       & 3         & 16        & 16               & softmax    & 64       & 2       & 64       & 0             & 0
        \\
        3   & 1.01     & 3       & 3         & 16        & 16               & softmax    & 32       & 2       & 64       & 0             & 0
        \\
        4   & 1.23     & 2       & 3         & 8         & 32               & exp        & 64       & 1       & 32       & 0             & 0
        \\
        5   & 1.74     & 2       & 2         & 8         & 8                & exp        & 32       & 1       & 32       & 0             & 0
        \\
        \bottomrule
    \end{tabular*}
\end{table}

\section{Speed Measurements}

\begin{table}[H]
    \begin{threeparttable}
        \caption{GPU scaling performance for a \num{15552}-atom polystyrene system (\qty{0.5}{\fs} time step) on varying numbers ($\ngpu$) of A100-SXM4 (\qty{80}{\giga\byte}) GPUs.
            While Vivace and MACE-OFF show strong scaling, the much faster PCFF is limited by communication overhead. The parallelization efficiency is governed by the model's cutoff radius and its computation-to-communication ratio. The MLFFs benefit from shorter cutoffs (MACE-OFF: 4.5 Å, Vivace: 6.5 Å) and higher computational intensity. In contrast, PCFF's longer cutoff (9--10 Å) and rapid calculation speed make communication the dominant bottleneck.
            \label{tab:speed_table}
        }
        \centering
        \setlength{\tabcolsep}{0pt}
        \begin{tabular*}{\linewidth}{@{}@{\extracolsep{\fill}}lS[table-format=2.2]S[table-format=2.2]S[table-format=2.2]S[table-format=2.2]@{}}
            \toprule
            & \multicolumn{4}{c}{$\ngpu$ (\unit{\nspday})}\\
            \cmidrule{2-5}
            Model & {$1$} & {$2$} & {$4$} & {$8$}
            \\
            \midrule
            PCFF &  10.24 & 11.08 & 10.71 & 8.22 \\
            Vivace &  0.30 & 0.55 & 0.80 & 1.18 \\
            MACE-OFF23-s & 0.29 & 0.35 & 0.87 & 1.20 \\
            UMA\tnote{\textdagger} & 0.02 & {-} & {-}  & {-} \\
            \bottomrule
        \end{tabular*}
        \begin{tablenotes}
            \item[\textdagger] Unable to test multi-GPU performance with iPi.
        \end{tablenotes}
    \end{threeparttable}
\end{table}

\begin{table}[H]
    \begin{threeparttable}
        \caption{Single GPU performance for varying system sizes (number of atoms $\natom$) of polystyrene at 1.0 g/cm$^3$ density. Performance was benchmarked on a single A100 (\qty{80} {\giga\byte}) GPU with a 0.5 fs timestep. All models are demonstrating a computational overhead, meaning that doubling the system size does not precisely halve the simulation speed.
            \label{tab:speed_per_atom}
        }
        \centering
        \setlength{\tabcolsep}{0pt}
        \begin{tabular*}{\linewidth}{@{}@{\extracolsep{\fill}}lS[table-format=2.2]S[table-format=2.2]S[table-format=2.2]S[table-format=2.2]@{}}
            \toprule
            & \multicolumn{4}{c}{$\natom$ (\unit{\nspday})}\\
            \cmidrule{2-5}
            Model & \num{1944} & \num{4050} & \num{7938} & \num{15552}
            \\
            \midrule
            PCFF & 19.40 & 18.98 & 17.68 & 11.28  \\
            Vivace & 1.22 & 0.84 & 0.52 & 0.27 \\
            MACE-OFF23-s &  1.41 & 0.86 &  0.51 & 0.27 \\
            \bottomrule
        \end{tabular*}
        \begin{tablenotes}
            \item[\textdagger] Unable to test multi-GPU performance with iPi.
        \end{tablenotes}
    \end{threeparttable}
\end{table}

\section{Additional Density and Glass Transition Figures}

\begin{figure}[H]
    \centering
    \includegraphics{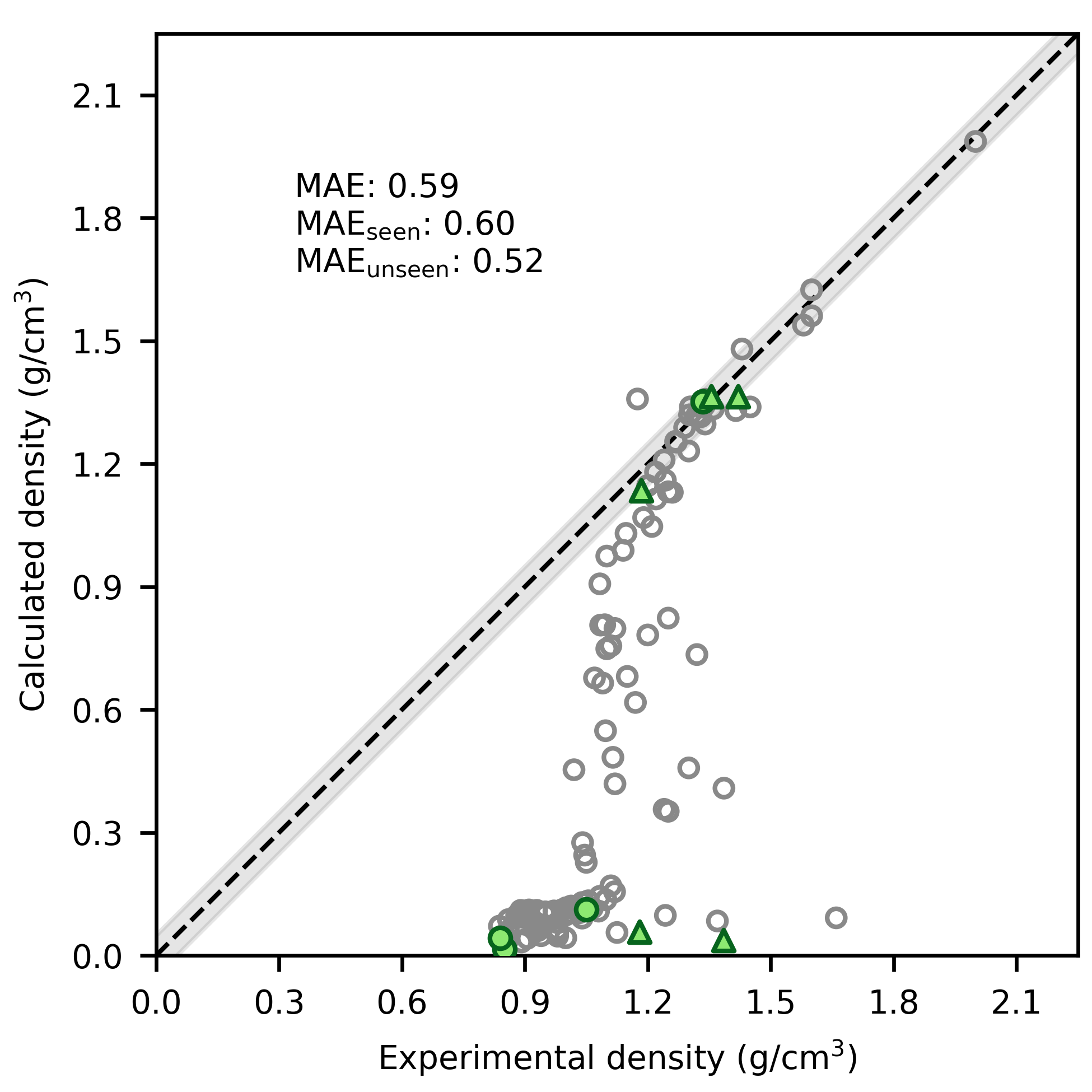}
    \caption{%
        Density parity plots for the pre-trained Vivace model before fine-tuning on PolyData.
        Circles ($\scalebox{0.8}{$\bigcirc$}$) and triangles ($\triangle$) indicate seen and unseen polymers respectively.
        Green markers indicate the set of 10 polymers investigated in more details in the main paper.
    }
    \label{fig:density_pretrained}
\end{figure}

\begin{figure}[H]
    \centering
    \includegraphics[width=\textwidth]{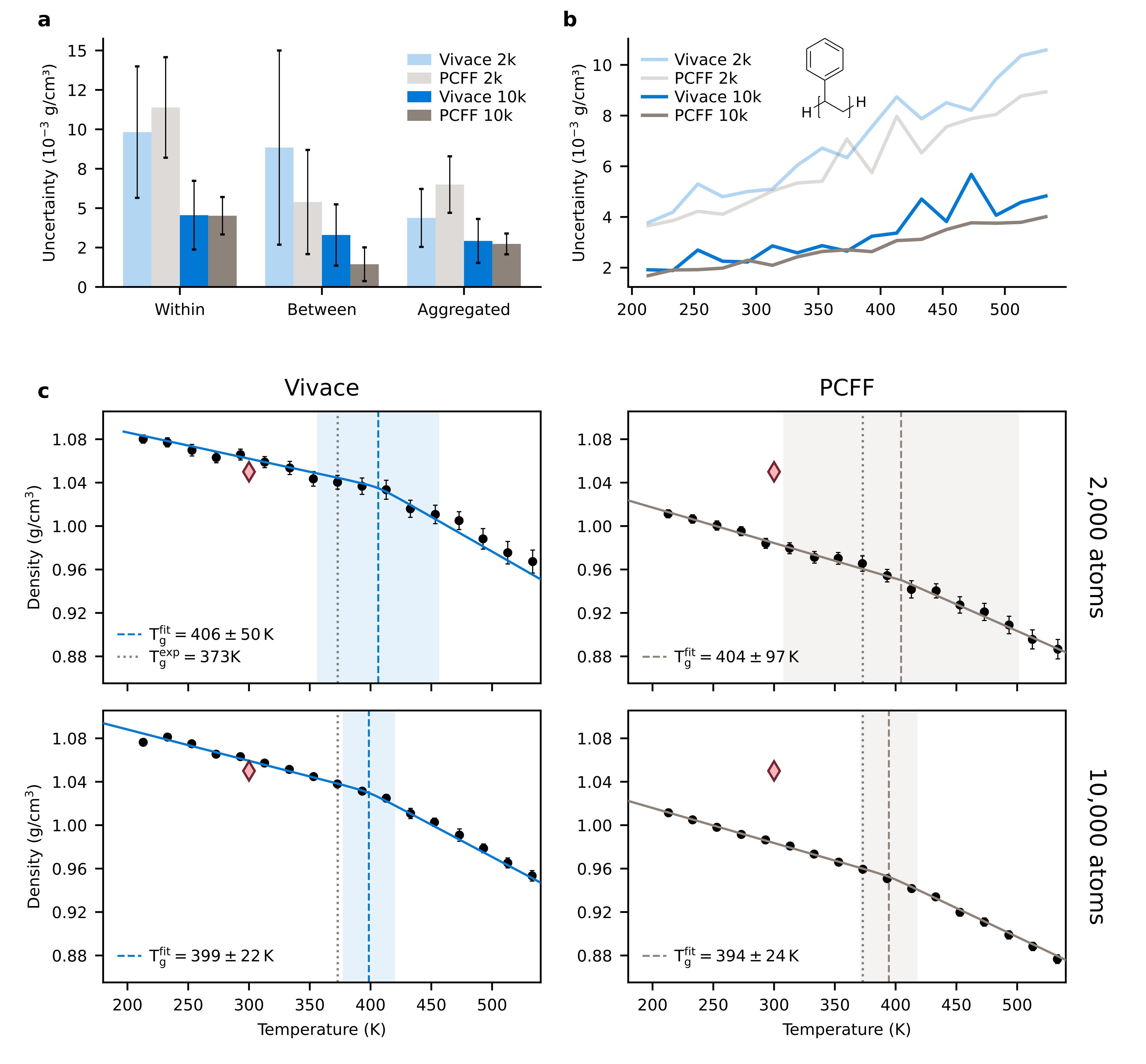}
    \caption{%
        Influence of the simulation system size on the derivation of density and \tg{} values.
        Given the variability in reported experimental values for densities, MD simulations containing approximately \num{2000} atoms can provide density estimates with acceptable precision.
        In contrast, it is crucial to use larger systems containing \num{10000} atoms in order to derive accurate \tg{} estimates.
        \textbf{a}, Mean density uncertainty within simulations, between simulations and in aggregated simulations (using weights inversely proportional to the within simulation uncertainty) for the 10 polymers in Table~\ref{tab:poly_ids}.
        \textbf{b}, Evolution with temperature of the density uncertainty in aggregated simulations for PS.
        \textbf{c}, Density vs temperature curves obtained for PS using Vivace (left) and PCFF (right).
        Top and bottom panels show results for different system sizes.
        Each point represents the average of three density simulations, with weights inversely proportional to the within simulation uncertainty.
        Error bars indicate the combined uncertainty, which is noticeably smaller at lower temperatures and when using \num{10000} atoms.
        Examples of fitted hyperbolas used to derive \tgfit{} are shown in blue.
        \tgfit{} and the experimental \tg{} are indicated by vertical dashed and dotted lines, respectively.
        The shaded area indicates the uncertainty of the fit, derived by bootstrapping.
        The red diamond indicates the experimental density at \qty{298.15}{\K}.
    }
    \label{fig:system_size}
\end{figure}

\begin{figure}[H]
    \centering
    \includegraphics[width=\textwidth]{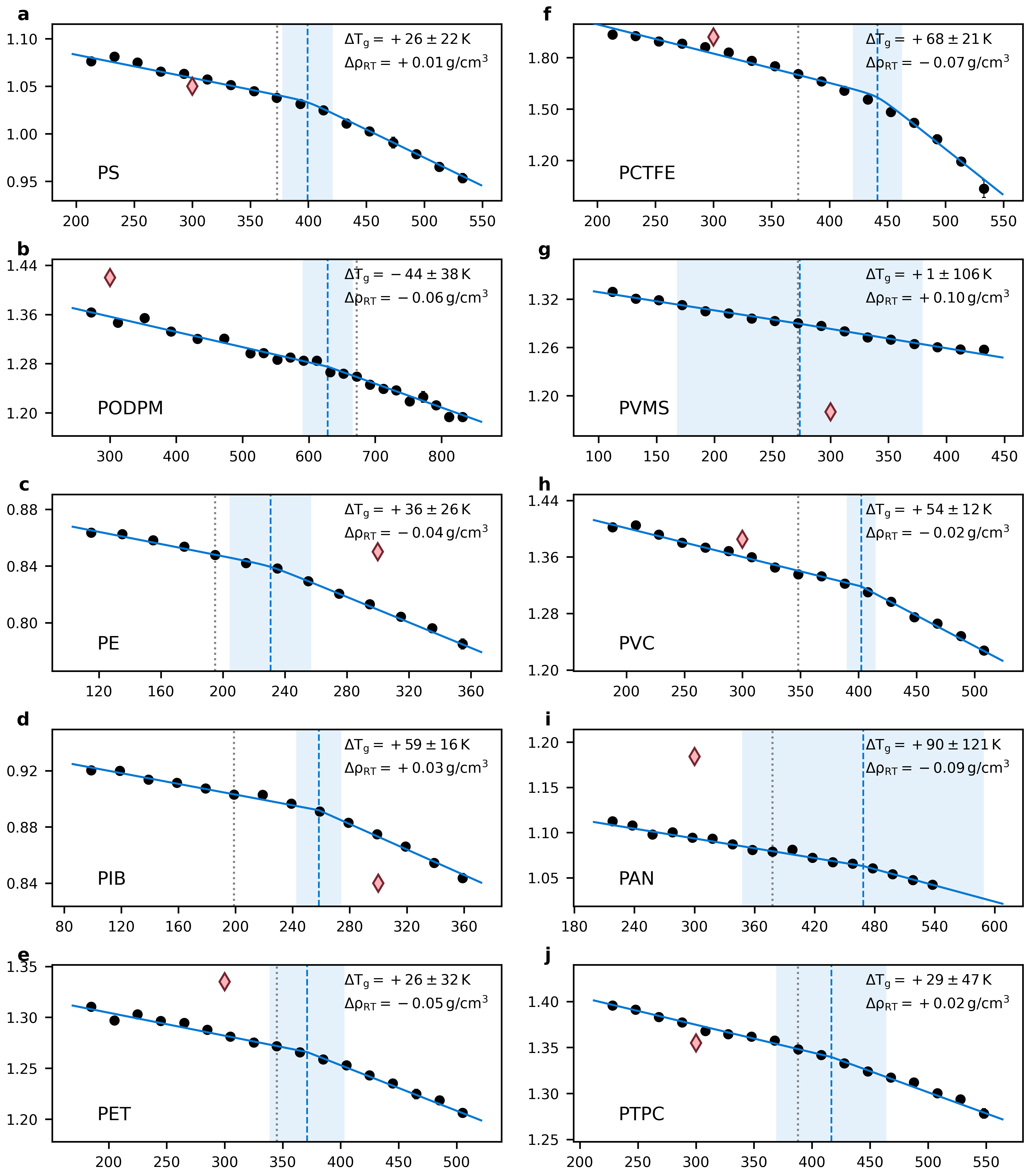}
    \caption{%
        Density vs temperature curves obtained with Vivace for 10 polymers selected from the PolyArena benchmark (see Table~\ref{tab:poly_ids} for abbreviations), with a hyperbola fit shown as a solid blue line. The \tg{} value derived by bootstrapping is shown as a dashed line and the shaded area represents the uncertainty of the estimate (see Section~\ref{sec:tg_sim} for details). The dotted line indicates the reference experimental value for \tg, while the red diamond symbol ($\lozenge$) indicates the experimental density at room temperature $\rho_{\mathrm{RT}}$.
    }
    \label{fig:tg_mlff}
\end{figure}

\begin{figure}[H]
    \centering
    \includegraphics[width=\textwidth]{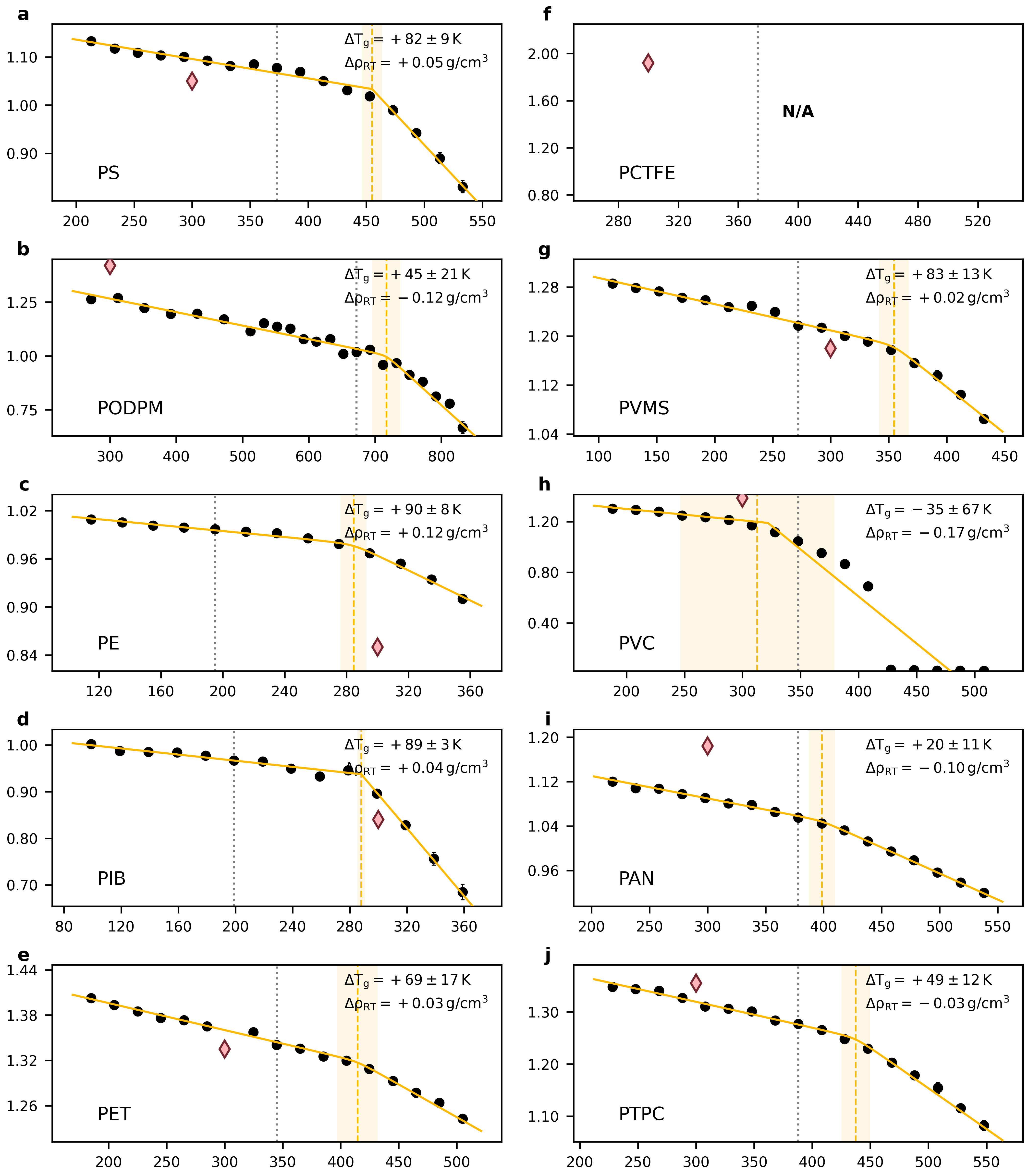}
    \caption{%
        Same as Figure~\ref{fig:tg_mlff}, but for MACE-OFF.
        N/A indicates polymers for which no \tg{} could be derived due to MD simulations failing.
        In the case of PVC we note simulations did not crash but became unstable with the density approaching 0 at higher temperatures.
    }
    \label{fig:tg_mace}
\end{figure}

\begin{figure}[H]
    \centering
    \includegraphics[width=\textwidth]{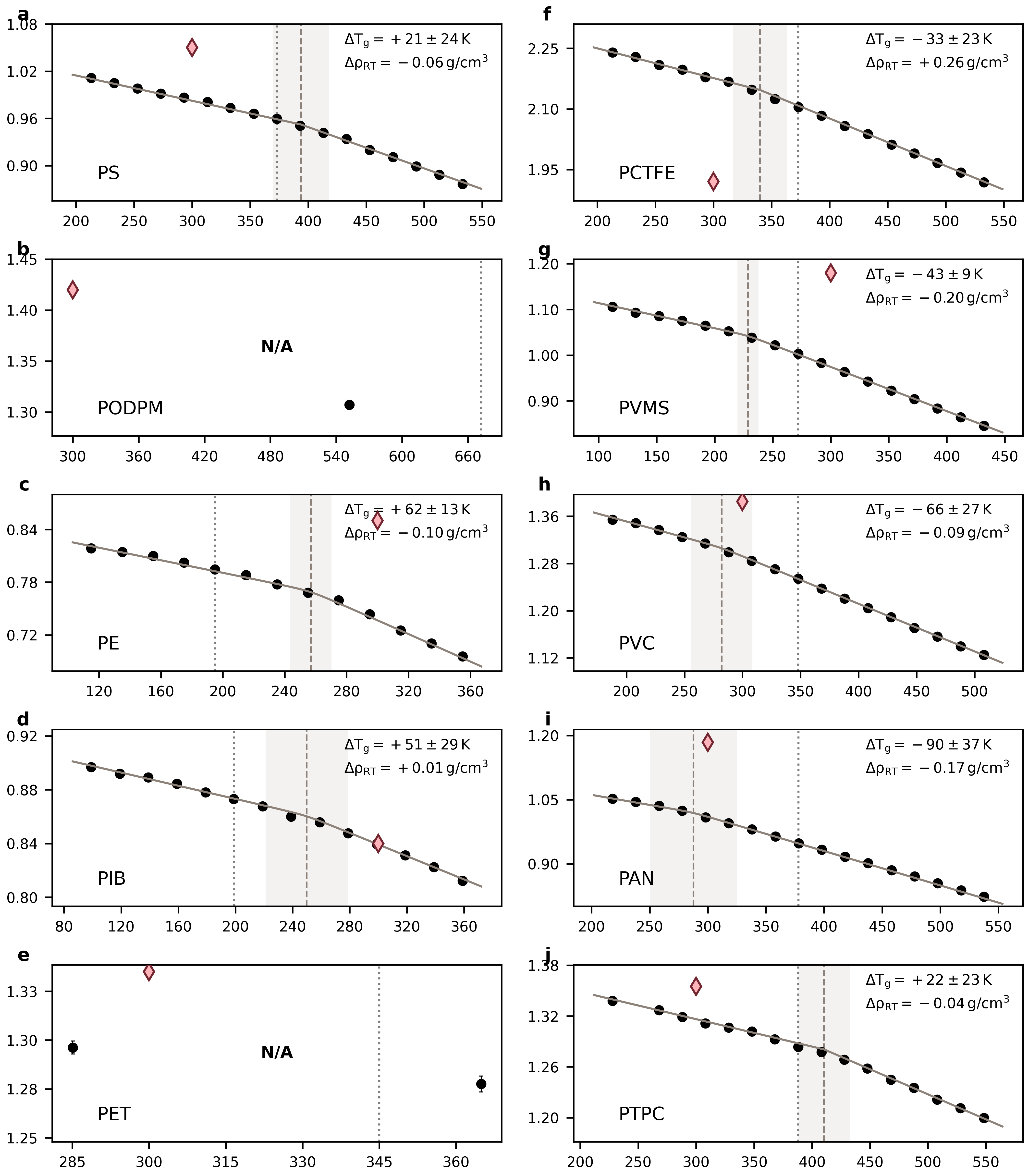}
    \caption{%
        Same as Figure~\ref{fig:tg_mlff}, but for PCFF. N/A indicates polymers for which no \tg{} could be derived due to MD simulations failing.
    }
    \label{fig:tg_pcff}
\end{figure}

\begin{figure}[H]
    \centering
    \includegraphics[width=\textwidth]{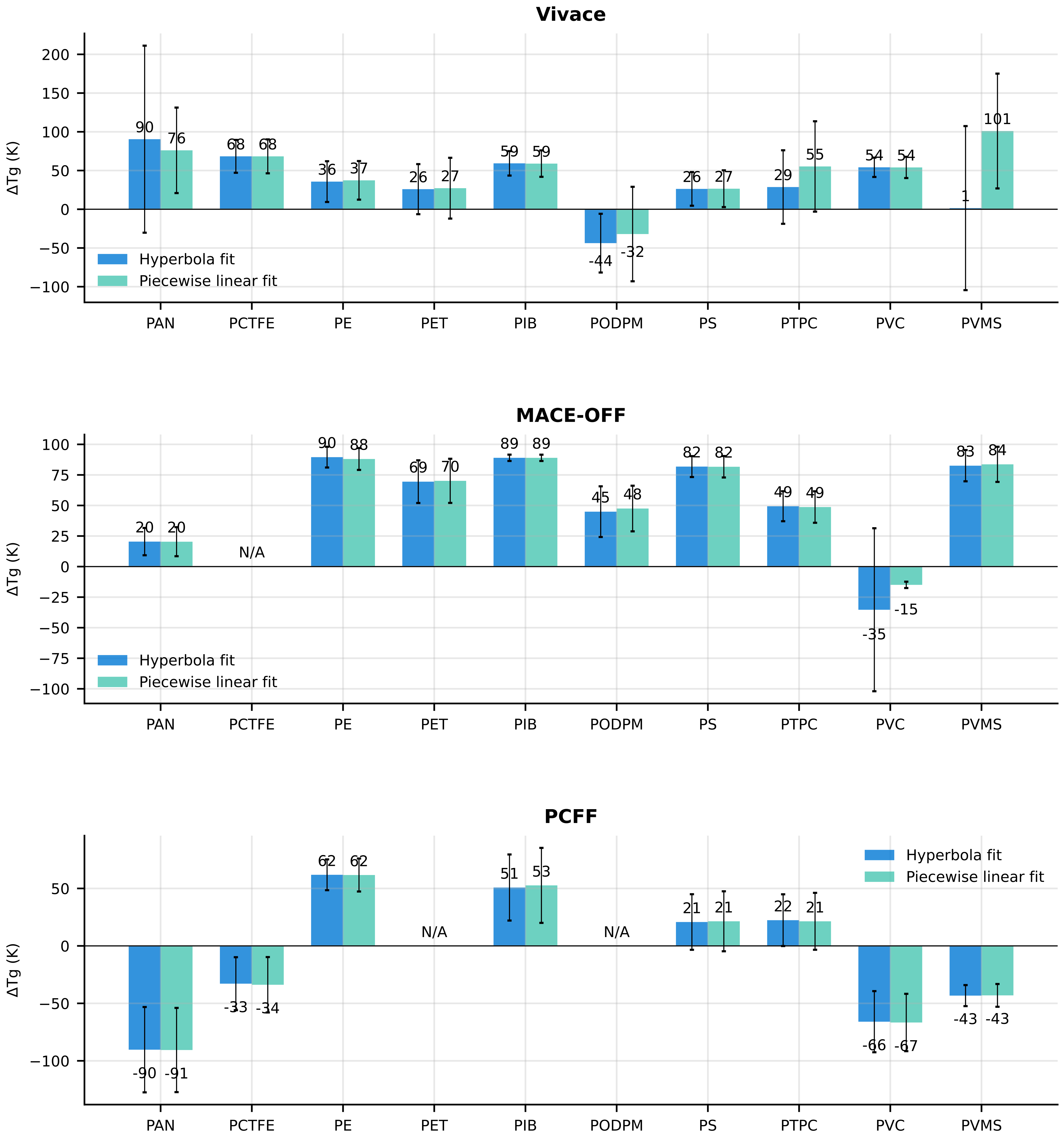}
    \caption{%
        Influence of the fitting method used to derive \tg.
        Fitting a hyperbola (blue), as described by \citeauthor{Patrone2016Uncertainty}\supercite{Patrone2016Uncertainty}, or a piecewise linear model (green) leads to nearly identical results.
        Importantly, both approaches make use of all the data available and do not require any user input.
        N/A indicates polymers for which no \tg{} could be derived due to MD simulations failing.
    }
    \label{fig:tg_methods}
\end{figure}

\printbibliography

\end{document}